\documentclass[aps,reprint,superscriptaddress,longbibliography,nofootinbib]{revtex4-2}
\usepackage{amssymb}
\usepackage{amsthm}
\usepackage{mathtools}
\usepackage{physics}
\usepackage{graphicx}
\usepackage{bm}
\usepackage[dvipsnames]{xcolor}
\usepackage{comment}

\newcommand{\noin}{}

\newtheorem{proposition}{Proposition}[section]

\newtheorem{theorem}{Theorem}[section]
\newtheorem{corollary}{Corollary}[section]
\newtheorem{lemma}{Lemma}[section]

\newcommand{\beq}{\begin{equation}}
\newcommand{\eeq}{\end{equation}}

\usepackage{hyperref}
\hypersetup{hidelinks}
\begin{document}
\title{Morphogenic necking of surfaces}
\author{Farzan Vafa}
\email{farzan.vafa@nbi.ku.dk}
\affiliation{Center of Mathematical Sciences and Applications, Harvard University, Cambridge, MA 02138, USA}
\affiliation{Physics of Living Systems, Department of Physics, Massachusetts Institute of Technology, Cambridge, Massachusetts 02139, USA}
\affiliation{Niels Bohr Institute, University of Copenhagen, 2100 Blegdamsvej 17, Copenhagen, Denmark}
\author{Puskar Mondal}
\email{pushkarmondal@gmail.com}
\affiliation{Center of Mathematical Sciences and Applications, Harvard University, Cambridge, MA 02138, USA}
\affiliation{Department of Mathematics, Harvard University, Cambridge, MA 02138, USA}
\affiliation{Beijing Institute of Mathematical Sciences and Applications, Tsinghua University, Beijing}
\author{L. Mahadevan}
\affiliation{Department of Mathematics, Harvard University, Cambridge, MA 02138, USA}
\affiliation{School of Engineering and Applied Sciences, Harvard University, Cambridge, MA 02138, USA}
\affiliation{Departments of Physics, and Organismic and Evolutionary Biology, Harvard University, Cambridge, MA 02138, USA}

\begin{abstract}
Surface morphogenesis occurs across biological scales, from subcellular vesicles and cells to visceral organs, tissue organoids, and whole organisms. Since morphogenesis often involves changes in topology, a natural question is that of appropriate mathematical descriptions that provide an approach to topological transitions of thin surfaces. Here, we suggest a general intrinsic framework coupling morphogen organization to surface geometry. We propose a minimal geometric model in which smoothing competes with morphogen-induced deformation to regulate shape, while loss of this balance can bring a surface toward a topological transition via neck constriction. A simple dynamical model coupling a scalar field to the surface metric reveals two mechanisms: gradient-driven constriction at steep morphogen transitions and curvature-driven constriction near localized morphogen peaks. For frozen morphogen profiles in simple axisymmetric examples, the gradient-driven route gives exact algebraic contraction, while the curvature-driven route gives local exponential contraction.
\end{abstract}
\maketitle
\section{Introduction}
\label{sec:introduction}

\begin{figure}[t]
\centering
\includegraphics[width=\columnwidth]{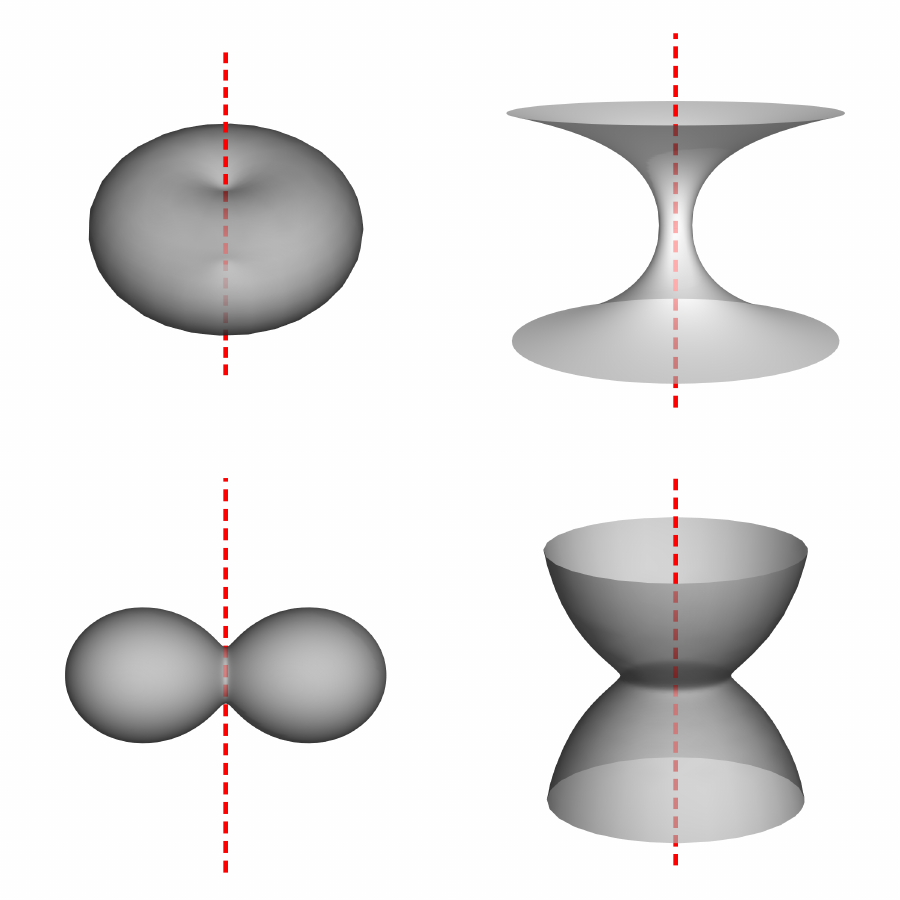}
\caption{\textbf{Schematic routes to topology change by pinching or fusion.} Top-left: the north and south poles of a sphere join to form a torus. Top-right: two disjoint planes merge along a tube to form a wormhole-like geometry (reverse process of pinching). Bottom-left: a topological sphere separates into two topological spheres. Bottom-right: a topological cylinder separates into two topological cylinders. In this paper, the calculations concern the local approach to constriction driven purely by intrinsic dynamics, and do not consider the global reconnection question.}
\label{fig:pinching}
\end{figure}

Morphogenesis is the process by which living systems acquire their shape and organization during development~\cite{barresi2020developmental}. Since early attempts to understand biological form through geometry and physical laws~\cite{thompson1917on}, it has become increasingly clear that morphogenesis involves not only changes in size and shape but also changes in topology, as structures divide, merge, or develop new connections and openings~\cite{thom1969topological,isaeva2014topology}. Understanding how such topological transitions arise from local physical and biochemical processes remains a fundamental problem in morphogenesis.

At the subcellular level, membrane remodeling during vesicle budding and scission involves the formation and collapse of narrow necks~\cite{hurley2010membrane}. Cell division separates a connected cellular domain into two through localized constriction, mediated by actomyosin rings in eukaryotes~\cite{pollard2010mechanics}, FtsZ-based division machinery in bacteria~\cite{adams2009bacterial}, and ESCRT assemblies in archaea~\cite{harker2022physical}. At larger scales, epithelial budding and tissue remodeling can involve changes in connectivity~\cite{ishihara2023topological}, often coordinated by spatially localized biochemical signals~\cite{munder2010notch}. Despite their diverse molecular origins, these processes share a geometric feature: localized deformation brings a structure toward a configuration in which its topology can change. Four examples are illustrated in Fig.~\ref{fig:pinching}. This suggests a common question: how can local biochemical organization drive geometric deformation toward a topological necking transition?

Most physical theories of biological shape change are formulated in terms of extrinsic geometry, which describes how a surface is embedded and bends in three-dimensional space. A familiar example is the Helfrich energy for fluid membranes, which combines mean and Gaussian curvature with a possible spontaneous curvature~\cite{Helfrich1973}. Such models have provided a framework for understanding vesicle budding, tubulation, and shape transitions through geometric phase diagrams~\cite{seifert1991shape}, and have been extended to experimentally controlled shape changes in epithelial organoids~\cite{ishihara2023topological}. When coupled to dissipation and fluid motion, they also describe the dynamics of membrane deformation~\cite{Seifert1999Dynamics}. However, topology is an intrinsic property, independent of how a surface is embedded. This motivates a complementary approach in which the primary dynamical variable is the surface metric, which determines distances, areas, and intrinsic curvature.

Biological geometry, however, does not evolve independently of biochemical organization. Spatially varying molecular signals can regulate growth, contraction, and remodeling, while the resulting geometric deformation can redistribute these signals or alter their activity. Such feedback has been explored in models coupling chemical pattern formation to surface deformation~\cite{Cummings1985,Mercker2013,Nesenberend2026}. Experiments demonstrate curvature-dependent recruitment of proteins such as annexin A4~\cite{Florentsen2021}, while theoretical studies show that coupling membrane composition to Gaussian curvature can destabilize narrow necks~\cite{ChenHiggsMacKintosh1997}. These observations suggest that biochemical organization and geometry should be treated as coupled dynamical fields. Here, we represent the local biochemical state by a scalar field, which we refer to broadly as a morphogen field~\cite{turing1952chemical}, coupled to the intrinsic surface metric.

A natural starting point for such a theory is provided by geometric flows. In two dimensions, Ricci flow acts as a form of geometric diffusion, redistributing Gaussian curvature and smoothing its spatial variations~\cite{hamilton,isenberg2011ricci}. But geometric smoothing alone need not produce the deformations required for topological change. For example, an ideal cylinder has zero Gaussian curvature and therefore remains stationary under unnormalized Ricci flow, irrespective of its radius. Localized constriction consequently requires an additional mechanism. Biochemical coupling provides one possibility: spatial variations in the morphogen field can drive metric deformation even where intrinsic curvature alone would produce no evolution. Conversely, mechanical and geometric relaxation can oppose such deformation, as occurs in biological tissues through processes that regulate growth, stress, and remodeling~\cite{guillot2013mechanics}. Together, these considerations motivate a minimal geometric model: biological form can be regulated by a balance between geometric smoothing and biochemically driven deformation, while the loss of this balance can lead to localized constriction and the approach to topological change.

Here we develop a minimal framework for intrinsic morphogenic geometrodynamics, in which an intrinsic surface metric is coupled to a scalar morphogen field through an effective energy. We focus on neck constriction as a geometric precursor to pinching and identify two distinct mechanisms for this process. The first is gradient-driven constriction, localized near steep spatial variations in the morphogen field. The second is curvature-driven constriction, arising from curvature coupling near localized morphogen maxima, where the morphogen gradient vanishes. By studying metric relaxation with a prescribed morphogen profile, we obtain an exact power-law solution and, under appropriate conditions, a local exponential contraction law in simple axisymmetric geometries. These results demonstrate how intrinsic geometric dynamics coupled to biochemical organization can drive a surface toward a singular configuration without prescribing an extrinsic constriction force. They describe the approach to pinching rather than the topological transition itself, whose completion requires additional physical mechanisms associated with scission, fusion, or tissue remodeling.

The paper is organized as follows. Section~\ref{sec:intrinsic-framework} develops the intrinsic energetic framework coupling geometry to the morphogen field. Section~\ref{sec:axisymmetric-dynamics} analyzes gradient-driven and curvature-driven neck constriction in axisymmetric geometries. Section~\ref{sec:general-dynamics} extends these examples to a general dynamical framework, and Section~\ref{sec:discussion} discusses their biological implications, limitations, and open questions. The appendices contain detailed derivations, stability analyses, mathematical estimates, energy balances, and alternative dynamical laws.

\section{A minimal intrinsic framework}
\label{sec:intrinsic-framework}

Let $(\Sigma,g)$ be a smooth $2-$ dimensional Riemann surface with its intrinsic metric $g_{ij}$ and denote its local morphogen organization by a scalar field $m$. Depending on the physical system, $m$ may represent a concentration, biochemical activity, or an effective order parameter; we leave its microscopic interpretation unspecified. We consider the energy functional
\begin{equation}
\mathcal F[m,g]
=\int_\Sigma d^2x\,\sqrt{g}
\left[
\frac K2 g^{ij}\nabla_i m\nabla_j m+V(R,m)
\right],
\label{eq:free-energy}
\end{equation}
where $K>0$ is the gradient stiffness, $R$ is the scalar curvature (twice the Gaussian curvature for surfaces), and $V(R,m)$ is a local potential coupling the morphogen to curvature.

\begin{figure}[t]
\centering
\includegraphics[width=\columnwidth]{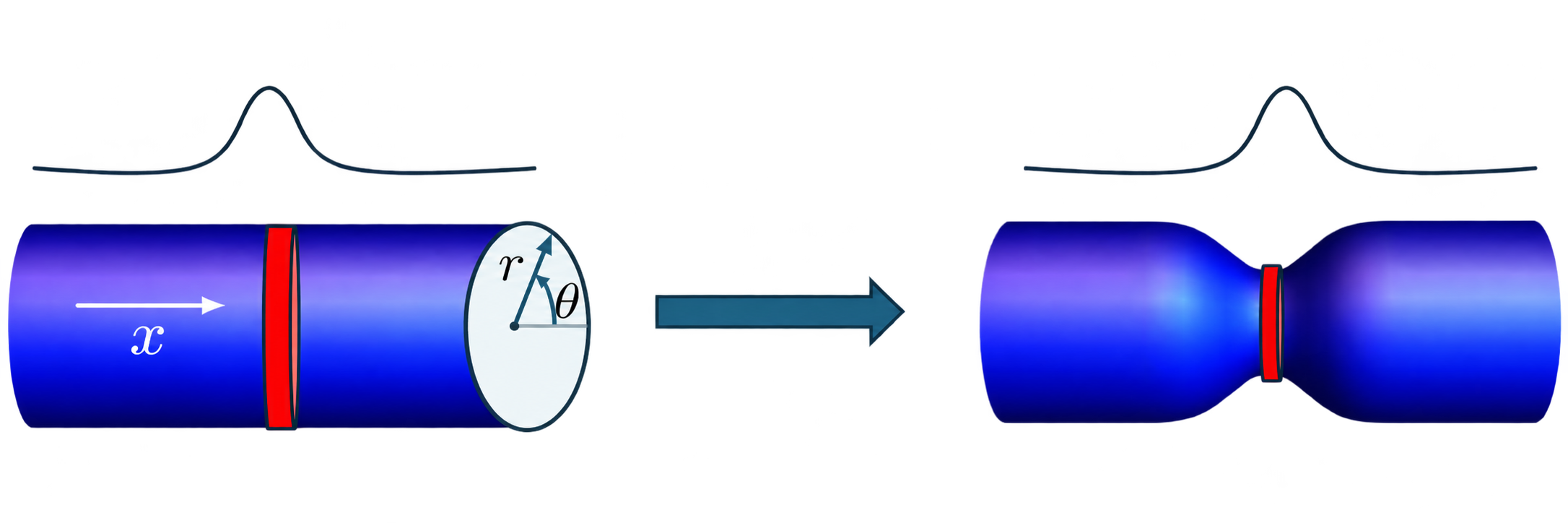}
\caption{\textbf{Schematic of morphogen-induced constriction in a cylindrical tube.} On the initial cylinder (left), $x$ labels points along the tube, $\theta$ is the azimuthal angle, and $r$ is the radius. Coupling to a localized morphogen pattern can favor transverse contraction, deforming the initial cylinder (left) into a neck (right). The upper curves represent either the scalar field $m$ or its squared gradient magnitude $|\nabla m|^2$; the red bands mark where the represented quantity is concentrated.}
\label{fig:pinching_example}
\end{figure}

\emph{Gradient energy.} At fixed geometry, the first term favors a uniform morphogen profile. Conversely, for a prescribed axial morphogen profile, a tube can lower this energy by contracting transversely. When local area is preserved, contraction is accompanied by axial stretching: the same change in $m$ then occurs over a larger physical distance. A morphogen profile that varies rapidly across a narrow band can therefore favor a neck there, even if the band is not centered on a maximum of $m$.

\emph{Curvature coupling.} To isolate the leading curvature-dependent contribution, we take
\beq V(R,m)=Rf(m), \label{eq:curvature-potential}\eeq 
keeping $f(m)$ general. The quadratic choice $f(m)\propto m^2$ has a particularly simple connection to the gradient energy: both terms are quadratic in the field and second order in spatial derivatives, with derivatives acting on $m$ in the gradient term and entering through the metric in $R$. In membranes, $Rf(m)$ represents a morphogen-dependent Gaussian rigidity, allowing local biochemical composition to alter the energetic cost of curvature~\cite{ChenHiggsMacKintosh1997}. Spatial variation of this rigidity can modify fusion pathways and their elastic barriers~\cite{Bottacchiari2024}. The quadratic form also arises naturally for tensor order parameters: covariant nematic elastic energies can generate a coupling $R\operatorname{tr}(Q^2)$, proportional to $Rm^2$ when $m$ denotes the magnitude of the nematic order tensor $Q$~\cite{vafa2022active}.

On a closed surface the integral of $R$ is fixed by Gauss--Bonnet, so a constant $f$ gives no local geometric force. A spatially varying $f(m)$ instead favors negative curvature in regions of larger $f$, with a compensating increase elsewhere. For a coupling increasing with $m$, a saddle-like neck can thus be favored near a peak of $m$, where the gradient term itself vanishes. Reversing the coupling, $f\to-f$, reverses this curvature preference and can instead favor positive curvature there; the gradient contribution is unchanged. The two terms therefore distinguish constriction at steep morphogen transitions from constriction near peaks (Fig.~\ref{fig:pinching_example}).

Scalar diffusion competes with these mechanisms by tending to smooth the spatial variations in $m$ that drive geometric deformation. Appreciable constriction therefore depends on whether the morphogen pattern persists long enough for the geometry to respond. Because deformation also changes transport along the surface, this competition evolves with the geometry.

We can quantify these effects for an initially cylindrical tube of radius $r_0$ (Fig.~\ref{fig:pinching_example}, left). Let $x$ measure its initial axial arclength and $r(x)$ denote the transverse radius after deformation. Holding $m(x)$ fixed, we vary $r$ while preserving local area and leaving the tube unchanged near its ends. With the coupling in Eq.~\eqref{eq:curvature-potential}, specializing Eq.~\eqref{eq:free-energy} to this axisymmetric geometry and integrating the curvature term by parts twice gives
\begin{equation}
\left.\frac{\delta\mathcal F}{\delta (r^2)} \right|_m
=\frac{\pi}{r_0}
\left[K(\partial_xm)^2-2\partial_x^2f(m)\right].
\label{eq:neck-variation-main}
\end{equation}
The energy is thus lowered wherever
\begin{equation}
K(\partial_xm)^2-2\partial_x^2f(m)>0.
\label{eq:pinching-criterion}
\end{equation}
In Eq.~\eqref{eq:pinching-criterion}, the first term (gradient) favors contraction, with the largest contribution at the steepest transitions. The second term (curvature coupling) favors contraction where $\partial_x^2f(m)<0$ and opposes it where this derivative is positive. Appendix~\ref{app:energetic-criterion} derives the energetic criterion. At a local maximum of $m$ with $\partial_x^2m<0$, the gradient contribution vanishes, while curvature coupling favors contraction if $f'(m)>0$. Whether this energetic tendency leads to sustained constriction depends on the coupled dynamics of the morphogen and geometry, which we consider next.

\subsection{From energy to dynamics}
The variations in energy determine which deformations lower $\mathcal F$, but their dynamics depend on the kinetics of $m$, transport, the lability of the geometry and the mechanical constraints on admissible deformations. With the curvature coupling introduced in Eq.~\eqref{eq:curvature-potential}, the intrinsic energy in Eq.~\eqref{eq:free-energy} becomes
\begin{equation}
\mathcal F[m,g]=\int_\Sigma dA_g\,
\left[\frac K2|\nabla m|_g^2+Rf(m)\right],\qquad K>0,
\label{eq:curvature-free-energy}
\end{equation}
on a smooth two-dimensional surface $\Sigma$.

We regard $m$ and $g_{ij}$ as independent configurational variables and assume a closed surface or variations compactly supported away from its boundary. Then, their conjugate variational responses are defined by
\begin{equation}
\mu=\frac{1}{\sqrt g}\left.\frac{\delta\mathcal F}{\delta m}\right|_g,
\qquad
\mathcal T^{ij}=\frac{2}{\sqrt g}
\left.\frac{\delta\mathcal F}{\delta g_{ij}}\right|_m,
\label{eq:variational-definitions}
\end{equation}
which satisfy $\delta\mathcal F=\int_\Sigma dA_g\, [\mu\,\delta m+\tfrac12\mathcal T^{ij}\delta g_{ij}]$. Here $\mathcal{T}^{ij}$ is the contravariant stress tensor associated with the pure geometric deformation. The generalized chemical potential $\mu$ measures the local energetic cost of changing $m$ at fixed geometry; in passive gradient flows, $-\mu$ drives nonconserved local relaxation, while gradients of $\mu$ drive diffusive transport toward lower chemical potential. When $m$ is an areal concentration and $\mathcal F$ is a thermodynamic free energy, $\mu$ has the usual chemical-potential interpretation.

Substituting the energy in Eq.~\eqref{eq:curvature-free-energy} into the variational definitions in Eq.~\eqref{eq:variational-definitions} gives
\begin{equation}
\mu=-K\Delta_gm+Rf'(m),
\label{eq:chemical-potential}
\end{equation}
\begin{align}
\mathcal T^{ij}={}&K\left(
\frac12g^{ij}|\nabla m|_g^2-\nabla^im\nabla^jm\right)\notag\\
&+2\left(\nabla^i\nabla^jf(m)-g^{ij}\Delta_gf(m)\right).
\label{eq:metric-response}
\end{align}
Appendix~\ref{app:variational-forces} derives these responses and their geometric interpretation; Appendix~\ref{app:density-variation} gives the alternative metric variation at fixed local amount for an areal concentration. We now see explicitly that the chemical potential has two contributions: gradient smoothing, through $-K\Delta_gm$, and curvature-dependent coupling, through $Rf'(m)$. For nonconserved passive relaxation, evolution proportional to $-\mu$ smooths variations in $m$, while intrinsic curvature can either enhance or oppose this relaxation.

To identify the geometric consequences of the metric stress, we decompose it into its traceless (deviatoric) and isotropic parts:
\begin{equation}
\mathcal T^{ij}
=\mathcal T_{\mathrm{dev}}^{ij}
+\mathcal T_{\mathrm{iso}}^{ij},
\qquad
\mathcal T_{\mathrm{iso}}^{ij}
=\frac12 g^{ij}\mathcal T^k{}_k.
\label{eq:stress-decomposition}
\end{equation}
In two dimensions, the gradient contribution is traceless, and the trace of the full stress is
\begin{equation}
\mathcal T^i{}_i=-2\Delta_g f(m).
\label{eq:stress-trace}
\end{equation}
Consequently,
\begin{subequations}\label{eq:stress-components}
\begin{align}
\mathcal T_{\mathrm{dev}}^{ij}
={}&K\left(
\frac12g^{ij}|\nabla m|_g^2
-\nabla^im\nabla^jm\right)
\nonumber\\
&+2\left(
\nabla^i\nabla^j f(m)
-\frac12g^{ij}\Delta_g f(m)\right),
\\
\mathcal T_{\mathrm{iso}}^{ij}
={}&-g^{ij}\Delta_g f(m).
\end{align}
\end{subequations}
This decomposition distinguishes directional deformation from local area change, although both contributions can arise from curvature coupling.

The gradient contribution to the deviatoric stress has principal values $-K|\nabla m|_g^2/2$ and $+K|\nabla m|_g^2/2$, parallel and perpendicular to $\nabla m$, respectively. Thus, longitudinal stretching or transverse contraction lowers the gradient energy, provided these deformations are mechanically admissible. Stretching reduces the physical gradient, while transverse contraction reduces the area over which it is distributed. Their combination favors longitudinal extension accompanied by transverse narrowing, providing one mechanism for neck constriction. The traceless character of this stress reflects the conformal invariance of the gradient energy in two dimensions: isotropic metric rescaling does not change the gradient energy.

Curvature coupling provides a second mechanism. Its isotropic stress is controlled by $-\Delta_g f(m)$, so that an infinitesimal isotropic contraction lowers the energy where $\Delta_g f(m)<0$, while expansion is favored where $\Delta_g f(m)>0$. At a smooth local maximum of $m$, where $\nabla m=0$ and $\Delta_g m\leq0$, we have
\begin{equation}
\Delta_g f(m)=f'(m)\Delta_gm.
\end{equation}
Thus, when $f'(m)>0$ and $\Delta_gm<0$, contraction is energetically favored even where the morphogen gradient vanishes. In addition, the traceless Hessian contribution in Eq.~\eqref{eq:stress-components} can drive anisotropic deformation near such extrema. Curvature coupling therefore generates both local area changes and directional deformation, providing a route to constriction distinct from the gradient-energy mechanism.

These energetic tendencies translate into geometric motion only after a kinetic law is specified. For example, under the simple dissipative relaxation
\begin{equation}
\partial_t g_{ij}=-M_g\mathcal T_{ij},
\label{eq:metric-gradient-relaxation}
\end{equation}
with $M_g>0$, the local area element evolves according to
\begin{equation}
\partial_t\log\sqrt g
=\frac12g^{ij}\partial_tg_{ij}
=M_g\Delta_g f(m).
\label{eq:local-area-relaxation}
\end{equation}
Hence the deviatoric stress changes the metric without directly changing the local area density, while the isotropic stress controls its instantaneous rate of change. Appendix~\ref{app:area-evolution} distinguishes local and total area balances.

The evolution of $m$ depends on whether it is conserved, transported, produced, or degraded, while metric evolution depends on the constitutive mobility and the geometric and mechanical constraints of the surface. Moreover, the minimal energy need not be bounded below under unrestricted metric variations, as discussed in Appendix~\ref{app:noncoercivity}. We therefore use these relaxation laws to identify mechanisms of geometric constriction.

\section[Two routes to neck constriction]{Two routes to neck constriction}
\label{sec:axisymmetric-dynamics}
To connect the energetic criterion in Eq.~\eqref{eq:pinching-criterion} to time-dependent constriction, we return to a tube geometry. Axisymmetry provides a simple setting in which a shrinking transverse circumference measures constriction, while the geometry and morphogen activity can still vary along the tube. Allowing longitudinal deformation captures an important feedback: stretching changes the physical morphogen gradient and transport along the surface, thereby modifying the contraction. We therefore describe the geometry using a transverse radius $r(x,t)$ and a longitudinal metric coefficient $h(x,t)$, writing the intrinsic metric as
\begin{equation}
ds^2=h(x,t)^2dx^2+r(x,t)^2d\theta^2 .
\label{eq:metric}
\end{equation}

For simplicity, we assume relaxational (gradient-flow) dynamics for both the nonconserved morphogen activity $m$ and the metric $g_{ij}$:
{
\beq
\partial_t m=-\Gamma_m\mu,\qquad
\partial_t g_{ij}=-\frac{\Gamma_g}{2}\mathcal T_{ij},
\label{eq:example-relaxation}
\eeq
}
\noin where $\Gamma_m,\Gamma_g\geq0$ are the scalar and geometric mobilities. Here $\mathcal T_{ij}=g_{ik}g_{j\ell}\mathcal T^{k\ell}$, with the energetic derivatives defined in Eq.~\eqref{eq:variational-definitions}. The metric law has the form of Eq.~\eqref{eq:metric-gradient-relaxation}, with $M_g=\Gamma_g/2$. We use fixed material coordinates, with metric variations taken at fixed $m$, and omit advection, sources, and Ricci flow. In the gradient-only case, we first allow the morphogen to diffuse and derive a general contraction identity. We then hold the morphogen profile fixed ($\Gamma_m=0$) to obtain explicit contraction laws for the two mechanisms. Appendix~\ref{app:relaxation-details} gives the relaxation laws and their energy balances.

\subsection{Exact gradient-driven contraction}
To isolate the effect of morphogen gradients on the geometry, we neglect curvature coupling ($f$ = 0). Model A then reduces to $\partial_tm=D_m\Delta_gm$, with scalar diffusivity $D_m \equiv K\Gamma_m\geq0$. In this gradient-only case, the stress is traceless [Eq.~\eqref{eq:stress-trace}], so the metric dynamics [Eq.~\eqref{eq:example-relaxation}] preserve local area. Transverse contraction is therefore accompanied by longitudinal stretching, with $hr=r_0$ throughout the evolution. Eliminating $h$ gives coupled evolution equations for the radius and morphogen activity:

\begin{subequations}\label{eq:tube_dynamics}
\begin{align}
\partial_tr
&=-\frac{\alpha}{8r_0^2}r^3(\partial_xm)^2,
\\
\partial_tm
&=\frac{D_m}{r_0^2}\partial_x
\left(r^2\partial_xm\right) ,
\end{align}
\end{subequations}
where $\alpha \equiv K\Gamma_g>0$ is the geometric reaction coefficient. Transverse circles thus contract wherever the morphogen gradient is nonzero, at a fractional rate proportional to the squared physical gradient $|\nabla m|_g^2=(r/r_0)^2(\partial_xm)^2$. Integrating the radius equation at a fixed material point gives

\begin{equation}
r(x,t)
=r_0\left[
1+\frac{\alpha}{4}
\int_0^t(\partial_xm(x,s))^2\,ds
\right]^{-1/2}.
\label{eq:exposure}
\end{equation}
This identity holds on every smooth interval of evolution, including when the morphogen diffuses. The radius records the accumulated squared coordinate gradient at each material point; reaching zero radius requires this integral to diverge. Appendix~\ref{sec:small-theorem} gives the precise collapse criterion.

For a frozen morphogen profile, $D_m=0$ and $m(x,t)=m(x,0)$, Eq.~\eqref{eq:exposure} reduces to
\begin{equation}
r(x,t)
=\frac{r_0}{
\sqrt{1+\frac{\alpha t}{4}
(\partial_xm(x,0))^2}}.
\label{eq:power_fit}
\end{equation}
Wherever the initial gradient is nonzero, the radius decays as $t^{-1/2}$ and $h$ grows as $t^{1/2}$ at late times. The stretching weakens the physical gradient $|\partial_xm|/h$ and therefore slows contraction, even though the morphogen profile is fixed in material coordinates. Appendix~\ref{app:frozen-gradient-solution} gives the corresponding full metric solution.

To obtain a localized neck, choose a smooth morphogen transition between two plateaus with Gaussian gradient
\begin{equation}
\partial_xm(x,0)
=\frac1a\exp\left(-\frac{x^2}{a^2}\right),
\qquad a>0.
\end{equation}
The minimum radius occurs at $x=0$:
\begin{equation}
r_{\min}(t)
=\frac{r_0}{
\sqrt{1+\alpha t/(4a^2)}}.
\label{eq:r_min}
\end{equation}
It approaches zero only at infinite time, but reaches any prescribed radius $\varepsilon r_0$, with $0<\varepsilon<1$, after a finite thinning time,

\begin{equation}
T_\varepsilon
=\frac{4a^2}{\alpha}
(\varepsilon^{-2}-1).
\label{eq:T}
\end{equation}
A microscopic cutoff can be reached in finite time even though the mathematical radius vanishes only at infinite time. Additional physics determines whether and how scission occurs at that scale. Appendix~\ref{app:gaussian-center-curvature} analyzes the curvature of this frozen Gaussian-gradient solution.

The initial region driving contraction has size of order $a$; its length along the surface grows as the tube narrows. If the morphogen is allowed to evolve, diffusion competes with contraction: smoothing can weaken the driving gradients. At the same time, meridional stretching reduces diffusion across fixed material labels. Whether a freely diffusing morphogen sustains contraction requires solving the coupled dynamics. Appendix~\ref{app:gradient-persistence} defines the initial width and derives local checks on the frozen-field approximation.

\subsection[Exponential curvature-driven contraction]{Exponential curvature-driven contraction}
We now retain the curvature coupling while keeping $m$ frozen ($\Gamma_m=0$), with the same full-metric law and initial cylinder, $h(x,0)=1$ and $r(x,0)=r_0$. At a local maximum of $m$ at $x=0$ with $\partial_x^2m(0)<0$, we have $\partial_xm(0)=0$. Both the gradient stress and $\partial_xf(m)$ vanish there, even for $K>0$. The meridional coefficient therefore remains fixed, $h(0,t)=1$, and the central radius obeys

\begin{equation}
\partial_t\ln r(0,t)
=\frac{\Gamma_g}{2}\left.\partial_x^2f(m)\right|_{x=0}.
\label{eq:exponential-origin}
\end{equation}
Since the morphogen profile is frozen, this fractional rate is constant, giving the exact law
\begin{equation}
r(0,t)=r_0\exp\!\left[
\frac{\Gamma_g t}{2}\left.\partial_x^2f(m)\right|_{x=0}
\right].
\label{eq:exponential-radius}
\end{equation}
At the peak, $\partial_x^2f(m)=f'(m)\partial_x^2m$, so the radius contracts exponentially when $f'(m)>0$. In the gradient-driven example, longitudinal stretching weakens the physical gradient and slows contraction. Here $h(0,t)$ stays fixed: the curvature coupling maintains a constant fractional contraction rate, and the local area decreases.

For example, if $f(m)=c m$ with $c>0$ and $m(x)=m_0\exp[-x^2/(2a^2)]$, with $m_0,a>0$, then $r(0,t)=r_0e^{-t/\tau_e}$, where $\tau_e=2a^2/(\Gamma_g c m_0)$.

This local law predicts a thinning time to radius $\varepsilon r_0$ of
\begin{equation}
T_\varepsilon=\tau_e\ln\frac1\varepsilon.
\label{eq:exponential-thinning-time}
\end{equation}
The exponential law holds at the peak for as long as the surrounding surface remains smooth. It does not determine the full neck shape or ensure that the peak remains the narrowest point. If smooth evolution persists for all time, the central radius approaches zero only as $t\to\infty$. Away from the peak, however, the frozen curvature-coupled dynamics amplify short-wavelength perturbations (Appendix~\ref{app:frozen-stability}).

For this Gaussian morphogen profile, the curvature contribution initially drives contraction over a region of width of order $a$. If the morphogen is allowed to relax, diffusion competes with contraction by tending to broaden and lower the peak, which can weaken the curvature-coupling drive. The morphogen also responds to surface curvature, so diffusion alone cannot determine whether the peak persists over the thinning time. Appendix~\ref{app:exponential-persistence} compares these effects and defines the initial driving width; the central exponential law does not determine the neck's later physical extent. These persistence estimates are formal: scalar diffusion alone does not remove the short-wavelength instability of this curvature-coupled law (Appendix~\ref{app:coupled-stability}).

The contraction laws describe intrinsic changes in the surface metric. Realizing this evolution as a surface of revolution in three-dimensional space requires the embedding condition derived in Appendix~\ref{app:embedding}. The frozen Gaussian-gradient solution eventually violates this condition. Whether the contraction mechanisms persist when embedding mechanics are included remains an open question.

\section[Extensions to generalized dynamics]{Extensions to generalized dynamics}
\label{sec:general-dynamics}

\begin{figure*}[t]
\centering
\includegraphics[width=\textwidth]{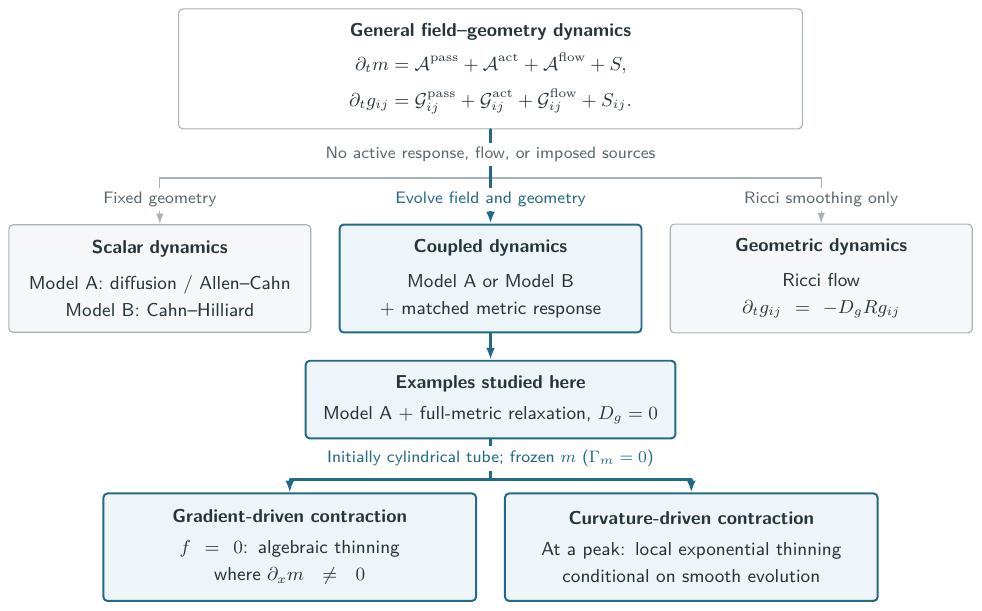}
\caption{\textbf{Representative evolution laws for morphogen fields and surface geometry.} The general dynamics in Eq.~\eqref{eq:general-dynamics} appear at the top, with passive, active, flow, and source contributions. Arrows indicate choices of response law and further specializations. The lower branches omit active responses, flow, and imposed sources. At fixed geometry, Model A (Eq.~\eqref{eq:nonconserved-option-main}) with $V=0$ gives diffusion; choosing $V=U(m)$ with a double-well potential gives Allen--Cahn (Model A) or Cahn--Hilliard (Model B, Eq.~\eqref{eq:density-option-main}) dynamics. In coupled passive relaxation, the metric law in Eq.~\eqref{eq:metric-option-main} uses $\mathcal T_{ij}$ for Model A and the density stress in Eq.~\eqref{eq:density-stress-main} for Model B under variations preserving local amount. Setting $\Gamma_g=0$ gives Ricci flow, which need not decrease $\mathcal F$. Blue paths select the examples studied here: the Model A and full-metric laws in Eq.~\eqref{eq:example-relaxation}, with $D_g=0$ and the curvature coupling in Eq.~\eqref{eq:curvature-potential}; the corresponding contraction laws are given in Eqs.~\eqref{eq:power_fit} and~\eqref{eq:exponential-radius}.
}
\label{fig:dynamics-map}
\end{figure*}

The preceding examples isolate two mechanisms of neck constriction under a specified law of geometric relaxation. Their explicit contraction laws assume a frozen morphogen profile; whether constriction persists when the morphogen evolves depends on the interplay between morphogen redistribution and geometric deformation. We now place these examples within a broader dynamical framework that includes passive relaxation, active driving, transport, and imposed sources. In a chosen coordinate frame, the coupled evolution of the scalar field and metric can be written as
\begin{subequations}\label{eq:general-dynamics}
\begin{align}
\partial_t m&=\mathcal A^{\mathrm{pass}}
+\mathcal A^{\mathrm{act}}+\mathcal A^{\mathrm{flow}}+S,
\label{eq:general-scalar}\\
\partial_t g_{ij}&=\mathcal G_{ij}^{\mathrm{pass}}
+\mathcal G_{ij}^{\mathrm{act}}+\mathcal G_{ij}^{\mathrm{flow}}+S_{ij}.
\label{eq:general-geometry}
\end{align}
\end{subequations}

The scalar rates $\mathcal A$ and symmetric tensor rates $\mathcal G_{ij}$ may depend on the fields, geometry, tangential velocity, and other system-specific variables. Passive terms describe energetic relaxation, active terms additional driving, and flow terms transport and coupling to motion. The separately specified sources $S$ and $S_{ij}$ represent scalar production, degradation, or exchange and imposed metric growth or remodeling, respectively; they are excluded from the constitutive rates. Active driving and imposed sources need not decrease $\mathcal F$.

The examples in Section~\ref{sec:axisymmetric-dynamics} use the passive Model A and full-metric laws in Eq.~\eqref{eq:example-relaxation}, with no advection, sources, or Ricci flow. Figure~\ref{fig:dynamics-map} maps representative scalar, geometric, and coupled evolution laws; blue paths identify the choices used in the examples. Appendix~\ref{app:transport-alternatives} details material transport and conservation laws. Appendix~\ref{app:general-transport-laws} gives the detailed alternatives, and Appendix~\ref{app:metric-energy} their energy balances. Appendices~\ref{app:ricci-flow} and~\ref{app:ricci-background} discuss Ricci flow and its geometric background. Nonnegative mobilities alone do not ensure a well-posed coupled evolution (Appendices~\ref{app:frozen-stability} and~\ref{app:coupled-stability}).

\section[Conclusions]{Conclusions}
\label{sec:discussion}
Coupling intrinsic geometry to a morphogen field provides a minimal framework for understanding how localized biochemical organization can drive geometric deformation. Our analysis identifies two distinct mechanisms of constriction: gradient-driven narrowing associated with spatial variations in morphogen activity, and curvature-coupled contraction that can occur even at localized morphogen maxima, where the gradient vanishes. The axisymmetric examples yield an exact power-law solution and, under appropriate conditions, a local exponential contraction law. While prescribed morphogen profiles isolate the geometric response to a fixed biochemical pattern, allowing the morphogen field to evolve in the gradient-only case shows that constriction depends on the accumulated history of its gradients. Together, these results demonstrate how local biochemical organization can drive intrinsic geometric evolution toward pinching without prescribing an extrinsic constriction force. They establish mechanisms of necking, but do not describe the completion of a topological transition.

An important next step is to examine the fully coupled dynamics of morphogen redistribution and geometric deformation. Localized biochemical patterns must persist long enough to induce appreciable constriction, while the evolving geometry may redistribute morphogens and modify their activity. Understanding this feedback could clarify the conditions under which localized patterns produce sustained constriction, transient deformation, or geometric instabilities.

Localized molecular assemblies involved in membrane constriction~\cite{harker2022physical}, cytokinesis~\cite{pollard2010mechanics}, and epithelial fusion~\cite{ishihara2023topological} provide opportunities for experimental tests. Time-resolved measurements of neck circumference $2\pi r(x,t)$, minimum radius $r_{\min}(t)$, and scalar curvature $R(x,t)$, together with morphogen activity $m(x,t)$, could distinguish constriction near sharp morphogen gradients from that occurring near morphogen maxima. Tracking material markers could further test the coupling between transverse narrowing and meridional extension predicted in the locally area-preserving example.

A broader theory must incorporate extrinsic geometry, mechanical stresses, and compatibility with embedding in three-dimensional space. As a neck approaches microscopic dimensions, additional physics determines whether and how scission, fusion, or tissue remodeling completes the topological transition. Finally, generalizing the scalar morphogen field to vector or tensor order parameters would allow orientational organization and active stresses to be included, as in tentacle formation in \emph{Hydra}~\cite{vafa2022active}. These extensions would connect the intrinsic geometric mechanisms identified here to the physical processes governing biological shape and topology.

\section*{Acknowledgments}

We thank Poul Martin Bendix, Tom Burkart, Erwin Frey, Alexander Grosberg, and Varun Venkatesh for helpful discussions. F.V. acknowledges support from the Novo Nordisk Foundation (grant No. NNF18SA0035142). This work was partially supported by the Center for Mathematical Sciences and Applications at Harvard University (F.V. and P.M.), where the project was initiated while both authors were based there. F.V. continued the work at MIT and is now based at the Niels Bohr Institute. P.M. is now based at the Beijing Institute of Mathematical Sciences and Applications. L.M. acknowledges partial support from the Simons Foundation and the Henri Seydoux Fund.

\section*{Contributions}

F.V.: conceptualization, formal analysis, investigation, writing---original draft, writing---review and editing; P.M.: conceptualization, formal analysis, investigation, writing---review and editing; L.M.: conceptualization, investigation, supervision, writing---original draft, writing---review and editing. All authors gave final approval for publication and agreed to be held accountable for the work performed therein.

\section*{Declaration of AI use}

We used ChatGPT for editorial assistance, checking analytical calculations, and preparing figures. The authors take full responsibility for the manuscript's content, results, and conclusions.

\bibliography{refs}

\appendix
% Preserve the numbering of the existing appendix equations and results.
\renewcommand{\theequation}{\thesection.\arabic{equation}}
\counterwithout{theorem}{section}
\counterwithout{proposition}{section}
\counterwithout{lemma}{section}
\counterwithout{corollary}{section}

\section[Energetic criterion for tube constriction]{Energetic criterion for tube constriction}
\label{app:energetic-criterion}
We derive Eq.~\eqref{eq:pinching-criterion} for the tube metric Eq.~\eqref{eq:metric}. The reference cylinder has $h=1$ and $r=r_0$, and $x$ measures its axial arclength. For an axisymmetric field $m=m(x)$,
\begin{equation}
|\nabla m|_g^2=\frac{(\partial_xm)^2}{h^2},
\qquad
R=-\frac{2}{hr}\partial_x
\left(\frac{\partial_xr}{h}\right).
\label{eq:axisymmetric-curvature}
\end{equation}

\noindent To compare the two energetic mechanisms, we hold $m(x)$ fixed and consider transverse contractions accompanied by meridional stretching, preserving local area:
\begin{equation}
hr=r_0,\qquad h=\frac{r_0}{r},
\label{eq:area-constraint}
\end{equation}
We restrict these deformations to coincide with the initial cylinder near the tube ends. Using $hr=r_0$, the curvature becomes
\begin{equation}
R=-\frac{1}{r_0^2}\partial_x^2(r^2).
\end{equation}
Substituting into Eq.~\eqref{eq:free-energy} with $V(R,m)=Rf(m)$ and integrating the curvature term twice by parts gives
\begin{equation}
\mathcal F
=\frac{\pi}{r_0}\int dx\,r^2
\left[
K(\partial_xm)^2
-2\partial_x^2f(m)
\right]
+B_{\partial},
\label{eq:neck-energy}
\end{equation}
where $B_{\partial}$ depends only on the endpoint data and is unchanged by the prescribed interior deformations. Consequently,
\begin{equation}
\delta\mathcal F
=\frac{\pi}{r_0}\int dx\,
\left[
K(\partial_xm)^2
-2\partial_x^2f(m)
\right]\delta(r^2).
\label{eq:neck-variation}
\end{equation}
A localized transverse contraction, $\delta(r^2)<0$, therefore lowers the energy wherever Eq.~\eqref{eq:pinching-criterion} holds. This is a criterion within the prescribed fixed-field, locally area-preserving trial family. The subsequent dynamics need not preserve local area when curvature coupling is present.

\noindent \emph{Gradient-driven constriction.---} When $f=0$, the coefficient in Eq.~\eqref{eq:pinching-criterion} is $K(\partial_xm)^2\geq0$. Hence contraction is energetically favored wherever the morphogen profile has a nonzero axial gradient. The accompanying increase in $h$ reduces the physical gradient $|\partial_xm|/h$, while the transverse circumference decreases. For equal local decreases in $r^2$, the energy reduction is greatest at the steepest morphogen transitions, where $|\partial_xm|$ is largest; the strongest drive need not occur at a peak of $m$.

\noindent \emph{Curvature-driven constriction.---} The second contribution in Eq.~\eqref{eq:pinching-criterion} favors contraction where $\partial_x^2f(m)<0$. Indeed,
\begin{equation}
\partial_x^2f(m)
=f''(m)(\partial_xm)^2
+f'(m)\partial_x^2m.
\label{eq:coupling-second-derivative}
\end{equation}
At a local maximum of $m$, where $\partial_xm=0$ and $\partial_x^2m<0$, this condition holds if $f'(m)>0$. Thus a curvature coupling increasing with the morphogen activity can favor constriction directly at its maximum, even though the gradient contribution vanishes there. If $f'(m)<0$, the curvature term instead favors transverse expansion at the maximum. Thus the two mechanisms can favor different locations within the localized activity profile illustrated in Fig.~\ref{fig:pinching_example}.

\noindent Within this trial family, continued narrowing lowers the energy wherever Eq.~\eqref{eq:pinching-criterion} holds, subject to geometric admissibility. A stable nonzero radius therefore requires additional mechanics or feedback. The actual evolution depends on the choice of dynamics.

\section[Variation and geometric interpretation]{Variation and geometric interpretation}
\label{app:variational-forces}
Let $\gamma_{ij}=\delta g_{ij}$, $\gamma=g^{ij}\gamma_{ij}$, and write $\Delta_g=\nabla^i\nabla_i$. All formulas below hold for compactly supported variations on the interior of a smooth surface, or for arbitrary smooth variations on a closed surface. The independent variations obey
\begin{equation}
\delta\sqrt g=\frac12\sqrt g\,\gamma,
\qquad
\delta g^{ij}=-g^{ik}g^{j\ell}\gamma_{k\ell}.
\label{eq:variation-metric-identities}
\end{equation}
Since $m$ is a scalar, $\delta(\nabla_i m)=\nabla_i(\delta m)$, and consequently
\begin{align}
\delta\mathcal F_{\rm grad}
={}&K\int_\Sigma dA_g\,\nabla^im\nabla_i(\delta m)
\nonumber\\
&+\frac K2\int_\Sigma dA_g\,
\left(\frac12g^{ij}|\nabla m|_g^2
-\nabla^im\nabla^jm\right)\gamma_{ij}.
\end{align}
Integration by parts of the first term gives $-K\Delta_gm$ as the field derivative and the displayed gradient contribution to $\mathcal T^{ij}$.

\noindent For the curvature term, the scalar-curvature variation is
\begin{equation}
\delta R=-R^{ij}\gamma_{ij}
+\nabla^i\nabla^j\gamma_{ij}-\Delta_g\gamma.
\label{eq:variation-scalar-curvature}
\end{equation}
In two dimensions $R_{ij}=\tfrac12Rg_{ij}$, so the terms proportional to $f(m)R$ cancel. Two integrations by parts then give
\begin{align}
\delta\!\int_\Sigma f(m)R\,dA_g
={}&\int_\Sigma Rf'(m)\,\delta m\,dA_g
\nonumber\\
&+\int_\Sigma
\bigl(\nabla^i\nabla^jf(m)
\nonumber\\
&\qquad-g^{ij}\Delta_gf(m)\bigr)
\gamma_{ij}\,dA_g.
\end{align}
Combining both variations proves Eqs.~\eqref{eq:chemical-potential} and \eqref{eq:metric-response}.

\noindent For an open surface and general boundary variations, the bulk first variation must be supplemented by
\begin{align}
\delta\mathcal F\big|_{\partial\Sigma}
={}&\int_{\partial\Sigma}ds_g\,\bigg[
K\,\delta m\,n^i\nabla_i m
\nonumber\\
&+f(m)n^i\left(\nabla^j\gamma_{ij}-\nabla_i\gamma\right)
\nonumber\\
&-n^j\gamma_{ij}\nabla^i f(m)
+\gamma\,n^i\nabla_i f(m)\bigg],
\label{eq:variation-boundary}
\end{align}
where $n$ is the outward unit conormal. Thus, on an open surface, spatial constancy of $f(m)$ removes the \emph{bulk} curvature force but does not in general remove boundary contributions.

\noindent The gradient contribution has zero trace in two dimensions, whereas
\begin{equation}
g_{ij}\mathcal T^{ij}=-2\Delta_g f(m).
\label{eq:metric-trace}
\end{equation}
At a point with $\nabla m\ne0$, choose an orthonormal frame with $e_\parallel=\nabla m/|\nabla m|_g$ and $e_\perp\perp e_\parallel$. In this frame the gradient part of $\mathcal T^i{}_j$ is
\begin{equation}
\frac{K|\nabla m|_g^2}{2}
\begin{pmatrix}-1&0\\0&1\end{pmatrix},
\end{equation}
and its local contribution to the first variation is
\begin{equation}
\frac K4|\nabla m|_g^2
\left[\delta g(e_\perp,e_\perp)
-\delta g(e_\parallel,e_\parallel)\right]dA_g.
\end{equation}
The sign describes favorable virtual metric variations, not a compulsory motion of an embedded elastic surface.

\noindent As a consistency check, diffeomorphism invariance of $\mathcal F$ gives the local identity
\begin{equation}
\nabla_i\mathcal T^{ij}=\mu\nabla^j m.
\label{eq:stress-noether}
\end{equation}
Indeed, substituting $\delta m=\xi^i\nabla_i m$ and $\delta g_{ij}=\nabla_i\xi_j+\nabla_j\xi_i$ into the first variation, and integrating by parts for any compactly supported vector field $\xi$, proves the identity. Equivalently, the curvature contribution satisfies $\nabla_i[2(\nabla^i\nabla^j f-g^{ij}\Delta_g f)] =2R^j{}_{k}\nabla^k f=R\nabla^j f$.

\noindent If $m$ instead denotes a conserved areal concentration, then the variation at fixed local material amount $\sqrt g\,m$ satisfies $\delta m=-\tfrac12m g^{ij}\delta g_{ij}$. The corresponding conjugate metric response is therefore
\begin{equation}
\mathcal T_{\rm dens}^{ij}
=\mathcal T^{ij}-m\mu g^{ij},
\label{eq:fixed-amount-stress}
\end{equation}
which must not be confused with the fixed-$m$ derivative in Eq.~\eqref{eq:variational-definitions}.

\noindent Finally, the minimal functional need not be bounded below under unrestricted metric variations. On a closed surface, let $f(m)$ be nonconstant and define the smooth conformal metrics $g_s=e^{-2s f(m)}g$, $s\ge0$. In two dimensions, the Dirichlet term is conformally invariant and $R_{g_s}dA_{g_s}=[R_g+2s\Delta_gf(m)]dA_g$, whence
\begin{equation}
\mathcal F[m,g_s]
=\mathcal F[m,g]
-2s\int_\Sigma|\nabla f(m)|_g^2\,dA_g
\longrightarrow-\infty.
\label{eq:noncoercivity}
\end{equation}
Thus this minimal energy identifies variational forces but cannot, without extra mechanical constraints or terms, establish stable geometric shape selection.

\section[Metric variation for a conserved density]{Metric variation for a conserved density}
\label{app:density-variation}
For an areal concentration, define the coordinate amount density $q=\sqrt g\,m$. If the local amount is held fixed under an instantaneous metric variation, then $\delta q=0$ yields
\begin{equation}
\delta m=-\frac m2g^{ij}\delta g_{ij}.
\end{equation}
The first-variation formula
\begin{equation}
\delta\mathcal F=\int_\Sigma\left(
\mu\,\delta m+\tfrac12\mathcal T^{ij}\delta g_{ij}
\right)dA
\end{equation}
therefore becomes
\begin{equation}
\left.\delta\mathcal F\right|_q
=\frac12\int_\Sigma
\left(\mathcal T^{ij}-m\mu g^{ij}\right)
\delta g_{ij}dA.
\label{eq:corrected-variation}
\end{equation}
This identifies $\mathcal T_{\mathrm B}^{ij}$ and makes explicit which local material quantity is fixed. Merely fixing the \emph{global} integral $\int mdA$ does not impose $\delta q(x)=0$ pointwise and does not, without a further constitutive specification, select this local stress.

\section[Local and total material area]{Local and total material area}
\label{app:area-evolution}
Throughout this section we set $v^i=0$, so that $\mathring g_{ij}=\partial_tg_{ij}$ and $\Theta=\partial_t\ln\sqrt g$. Since $\Theta=\tfrac12g^{ij}\mathring g_{ij}$, applying Eq.~\eqref{eq:metric-option-main} gives
\begin{equation}
\Theta=-\frac{\Gamma_g}{4}\mathcal T^i{}_i-D_gR.
\end{equation}
The special two-dimensional trace identity is
\begin{equation}
\mathcal T^i{}_i=-2\Delta_g f(m),
\label{eq:trace}
\end{equation}
where the Dirichlet contribution vanishes identically. Consequently,
\begin{equation}
\Theta=\frac{\Gamma_g}{2}\Delta_g f(m)-D_gR.
\end{equation}
For the complete closed surface, the transport formula gives
\begin{eqnarray}
\frac{d}{dt}\operatorname{Area}_g(\Sigma)
&=\int_\Sigma\Theta dA\nonumber\\
&=-D_g\int_\Sigma RdA
=-4\pi D_g\chi(\Sigma).
\label{eq:gauss-bonnet-area}
\end{eqnarray}
The vanishing integral of $\Delta_g f$ requires the closed surface assumption; on an open tube it is a boundary flux. For $D_g=0$ the total area is preserved, but $\Theta=\Gamma_g\Delta_gf/2$ need not vanish pointwise. When $f$ is constant, the same law preserves local area. A globally area-preserving alternative for Model A on a closed surface is to replace $-D_gRg_{ij}$ by $-D_g(R-\overline R)g_{ij}$, with $\overline R=\operatorname{Area}^{-1}\int RdA$. This modification is not local area conservation.

\noindent If Model B uses the correct fixed-$q$ stress, then
\begin{equation}
\Theta_{\mathrm B}
=\frac{\Gamma_g}{2}\bigl(\Delta_gf+m\mu\bigr)-D_gR.
\label{eq:area-B}
\end{equation}
Thus total-area conservation is not automatic even when $D_g=0$. A local area constraint requires $\Theta=0$ and an appropriate constrained metric law; equivalently, for a purely metric material rate one may project the prescribed rate to its traceless part. In two dimensions such a local projection removes a purely conformal Ricci-flow term.

\section[Noncoercivity of the minimal functional]{Noncoercivity of the minimal functional}
\label{app:noncoercivity}
On a closed surface, fix a smooth $m$ such that $f(m)$ is nonconstant, and consider the conformal metrics $g_s=e^{-2sf(m)}g$, $s\ge0$. In two dimensions the Dirichlet energy is conformally invariant, while for $\widetilde g=e^{2u}g$ one has $R_{\widetilde g}dA_{\widetilde g}=(R_g-2\Delta_gu)dA_g$. Two integrations by parts give
\begin{equation}
\mathcal F[m,g_s]
=\mathcal F[m,g]
-2s\int_\Sigma|\nabla f(m)|_g^2 dA.
\end{equation}
Thus $\mathcal F$ is unbounded below under unrestricted metric variations whenever $f(m)$ is nonconstant. Formal energy dissipation does not by itself guarantee a stable minimum or prevent metric degeneration. Additional mechanical costs or geometric constraints are required for a globally stable physical model.

\section[Details of the relaxation laws]{Details of the relaxation laws}
\label{app:relaxation-details}
\subsection{Area evolution}
\label{app:relaxation-area-evolution}
Neither fixed-$h$ radius relaxation nor full-metric relaxation imposes area conservation in general. Taking the trace of Eq.~\eqref{eq:metric-option-main} gives
\begin{equation}
\partial_t\ln\sqrt g
=\frac{\Gamma_g}{2}\Delta_gf(m)-D_gR.
\label{eq:area-evolution}
\end{equation}
Fixed local or total area therefore requires an appropriate constraint, unless the chosen terms already preserve it.

\noindent In the gradient-only example, $f=D_g=0$, the gradient stress is traceless in two dimensions, giving
\begin{equation}
\partial_t\ln\sqrt g
=\frac12g^{ij}\partial_tg_{ij}=0,
\end{equation}
so local area is preserved. For the initial cylinder $h=1$, $r=r_0$, this yields $hr=r_0$ throughout smooth evolution, as used in Eq.~\eqref{eq:tube_dynamics}.

\subsection{Full-metric energy balance}
\label{app:full-metric-energy}
For the full-metric law on a closed surface, the metric contribution is
\begin{equation}
\left.\frac{d\mathcal F}{dt}\right|_{m}
=
-\frac{\Gamma_g}{4}
\int_\Sigma|\mathcal T|_g^2\,dA_g
+D_g\int_\Sigma R\Delta_gf(m)\,dA_g.
\label{eq:metric-dissipation}
\end{equation}
The variational contribution is nonpositive. The Ricci contribution has no definite sign, so the combined law need not lower $\mathcal F$. In the pure-gradient, zero-Ricci Model A example, the complete passive energy balance is
\begin{equation}
\frac{d\mathcal F}{dt}
=-\Gamma_m\int_\Sigma\mu^2\,dA_g
-\frac{\Gamma_g}{4}
\int_\Sigma|\mathcal T|_g^2\,dA_g
\leq0.
\label{eq:analytic-dissipation}
\end{equation}
This identity holds on a closed surface or under boundary conditions eliminating energy fluxes. Scalar relaxation and metric deformation then dissipate the same functional, at rates set by their independent mobilities.

\subsection{Radius relaxation at fixed \texorpdfstring{$h$}{h}}
\label{app:radius-relaxation}
The fixed-$h$ radius law, Eq.~\eqref{eq:radius-smoothing-main}, combines variational relaxation with Ricci-inspired smoothing. We first give its purely variational part, obtained by setting $D_g=0$. In fixed material coordinates, without advection or sources, this law is
\begin{equation}
\partial_tr
=-\frac{\Gamma_r}{2\pi hr}
\left.
\frac{\delta\mathcal F}{\delta r(x)}
\right|_{h,m},
\qquad \partial_th=0.
\label{eq:radius-relaxation}
\end{equation}
Here $\Gamma_r\geq0$ is the radius mobility, and the functional has been integrated over the azimuthal angle.

\subsection{Radius force, smoothing, and energy balance}
\label{app:radius-energy}
For the axisymmetric functional, varying the radius at fixed $h$ and $m$, with variations supported away from the tube ends, gives
\begin{equation}
\left.
\frac{\delta\mathcal F}{\delta r(x)}
\right|_{h,m}
=
\frac{\pi K}{h}(\partial_xm)^2
-4\pi\partial_x
\left(\frac{\partial_xf(m)}{h}\right).
\label{eq:radius-force}
\end{equation}
The factor $1/(2\pi hr)$ in Eq.~\eqref{eq:radius-relaxation} divides this force by the local surface-area density. At fixed $h$ and $m$, the resulting energy rate is
\begin{equation}
\left.\frac{d\mathcal F}{dt}\right|_{h,m}
=
-\Gamma_r\int dx\,
\frac{1}{2\pi hr}
\left(
\frac{\delta\mathcal F}{\delta r}
\right)^2
\leq0,
\label{eq:radius-dissipation}
\end{equation}
provided $h,r>0$ and the boundary variations vanish.

\noindent For the diagonal axisymmetric metric, the scalar curvature is
\begin{equation}
R=-\frac{2}{hr}\partial_x\!\left(\frac{\partial_xr}{h}\right).
\label{eq:radius-smoothing-curvature}
\end{equation}
The radius component of Ricci flow thus gives $\partial_tr=-(D_g/2)Rr=(D_g/h)\partial_x[(\partial_xr)/h]$. Retaining this component with fixed $h$ gives the smoothing term in Eq.~\eqref{eq:radius-smoothing-main}. Writing $Q=\left.\delta\mathcal F/\delta r(x)\right|_{h,m}$, its combined energy rate, under the same boundary conditions, is
\begin{equation}
\begin{aligned}
\left.\frac{d\mathcal F}{dt}\right|_{h,m}
={}&-\Gamma_r\int\frac{Q^2}{2\pi hr}\,dx\\
&+D_g\int\frac{Q}{h}\partial_x\!\left(\frac{\partial_xr}{h}\right)dx.
\end{aligned}
\label{eq:radius-smoothing-dissipation}
\end{equation}
The added term has no definite sign for $\mathcal F$; the nonpositive identity in Eq.~\eqref{eq:radius-dissipation} applies to the purely variational law.
\section[Estimates for the coupled morphogen--metric flow]{Estimates for the coupled morphogen--metric flow}
\label{app:estimates}
We write spatial derivatives using $\partial_x$ in the \emph{fixed} spatial coordinate $x$; this coordinate is generally not arclength at positive times. Our convention is $\Delta=\operatorname{div}_g\nabla$, and $R$ is the scalar curvature, twice the Gaussian curvature. The correspondence with the rates in the main text is
\begin{eqnarray}
 \Gamma_g=\gamma_g^{-1}, \Gamma_ m=\gamma_ m^{-1},
 \alpha=\frac{K}{\gamma_g}, D_ m=\frac{K}{\gamma_ m},\\
 \delta=\frac{\gamma_g}{\gamma_ m}=\frac{D_ m}{\alpha}.
 \label{app:parameters}
\end{eqnarray}
Thus $\delta\ll1$ means that morphogen diffusion is slow relative to the metric response, whereas $\delta\gg1$ means that it is fast. The estimates below hold for all finite positive $K,\gamma_g,\gamma_m$ without assuming a separation of timescales.

\subsection{Gauge Invariant identities and the axisymmetric reduction}
We first record the invariant equations, allowing $D_g\geq0$:
\begin{align}
 \partial_t g_{ij}
 &=\frac{\alpha}{2}\left(\nabla_i m\nabla_j m
       -\frac12|\nabla m|_g^2g_{ij}\right)-D_g Rg_{ij},
 \label{app:general-metric}\\
 \partial_t m&=D_ m\Delta m.
 \label{app:general-rho}
\end{align}
There is no potential term in this system.

\begin{proposition}[Scalar curvature and Bochner identities]
\label{app:invariant-identities}
On any smooth existence interval of \eqref{app:general-metric}--\eqref{app:general-rho},
\begin{align}
 \partial_t d\mu_g&=-D_gR\,d\mu_g,\label{app:volume-general}\\
 \partial_t R
 &=\frac{\alpha}{2}\operatorname{div}_g((\Delta m)\nabla m)\notag\\
 &\quad +D_g(\Delta R+R^2),\label{app:scalar-general}\\
 (\partial_t-D_ m\Delta)|\nabla m|_g^2
 &=-2D_ m|\nabla^2 m|_g^2\notag\\
 &\quad +(D_g-D_ m)R|\nabla m|_g^2
       \\
       &-\frac{\alpha}{4}|\nabla m|_g^4.
 \label{app:bochner}
\end{align}
\end{proposition}

\begin{proof}
In dimension two, $\operatorname{Ric}_{ij}=\frac12Rg_{ij}$. The morphogen stress in Eq.~\eqref{app:general-metric} is trace free, and hence $g^{ij}\partial_tg_{ij}=-2D_gR$, proving Eq.~ \eqref{app:volume-general}. For an arbitrary variation of the metric,
\begin{equation}
 \partial_tR
 =-\langle\operatorname{Ric},\partial_tg\rangle_g
     +\nabla^i\nabla^j(\partial_tg_{ij})
     -\Delta(g^{ij}\partial_tg_{ij}).
 \label{app:variation}
\end{equation}
Metric compatibility and the symmetry of the scalar Hessian give
\begin{align*}
 &\nabla^i\left(\nabla_i m\nabla_j m
              -\tfrac12|\nabla m|_g^2g_{ij}\right)
 \\&= (\Delta m)\nabla_j m
       +\nabla^i m\nabla_i\nabla_j m
       -\tfrac12\nabla_j|\nabla m|_g^2\\
 &= (\Delta m)\nabla_j m.
\end{align*}
Its contraction with $\operatorname{Ric}$ vanishes. The contribution of $-D_gRg$ to the three terms in Eq.~\eqref{app:variation} is, respectively, $D_gR^2$, $-D_g\Delta R$, and $2D_g\Delta R$. This proves Eq.~\eqref{app:scalar-general}.

\noindent Differentiating the inverse metric as well as $ m$, we obtain
\begin{align*}
 \partial_t|\nabla m|_g^2
 &=-\frac{\alpha}{4}|\nabla m|_g^4
       +D_gR|\nabla m|_g^2\\
 &\quad +2D_ m\langle\nabla m,\nabla\Delta m\rangle_g.
\end{align*}
The identity
\[
 \Delta|\nabla m|_g^2
 =2|\nabla^2 m|_g^2
   +2\langle\nabla m,\nabla\Delta m\rangle_g
   +R|\nabla m|_g^2
\]
then gives Eq.~\eqref{app:bochner}. In particular, the changing metric contributes the quartic damping term; a fixed-metric heat estimate cannot be substituted for Eq.~\eqref{app:bochner}.
\end{proof}

\noindent For the remainder of the appendix we set $D_g=0$ and impose azimuthal symmetry. The equations become
\begin{align}
 \partial_tg_{xx}&=\frac{K}{4\gamma_g}(\partial_x m)^2,
 \label{eq:gxx}\\
 \partial_tg_{\theta\theta}
 &=-\frac{K}{4\gamma_g}\frac{g_{\theta\theta}}{g_{xx}}(\partial_x m)^2,
 \label{eq:gtt}\\
 \partial_t m
 &=\frac{K}{\gamma_ m g_{xx}}
   \left( (\partial_x^2 m)+\frac12\partial_x
                  \log\frac{g_{\theta\theta}}{g_{xx}}\, (\partial_x m)\right).
 \label{eq:rho}
\end{align}
\par

\begin{lemma}[Conservation of the volume density]
\label{product}
For every fixed $x$,
\begin{equation}
 g_{xx}(t,x)g_{\theta\theta}(t,x)
 =g_{xx}(0,x)g_{\theta\theta}(0,x)=:C(x).
 \label{app:product}
\end{equation}
In particular $g_{xx}$ is nondecreasing and $g_{\theta\theta}$ is nonincreasing as long as the solution is smooth.
\end{lemma}

\begin{proof}
The product rule and Eqs.~\eqref{eq:gxx}--\eqref{eq:gtt} give
\[
 \partial_t(g_{xx}g_{\theta\theta})
 =\frac{K}{4\gamma_g}(\partial_x m)^2g_{\theta\theta}
  -\frac{K}{4\gamma_g}g_{\theta\theta}(\partial_x m)^2=0.
\]
The monotonicity statements follow from the signs in the two equations.
\end{proof}

\noindent Conservation in time does not imply that $C(x)$ is spatially constant. In all subsequent coordinate identities and estimates we assume
\begin{equation}
 C(x)\equiv C>0.
 \label{app:constant-density}
\end{equation}
This holds for the initial cylinder $g_{xx}(0,x)=1$, $g_{\theta\theta}(0,x)=r(0)^2$ used in the main text. Under Eq.~\eqref{app:constant-density},
\begin{equation}
 g_{\theta\theta}=\frac{C}{g_{xx}},\qquad
 \partial_t m=\frac{K}{\gamma_ m}
                    \partial_x\left(\frac{ (\partial_x m)}{g_{xx}}\right).
 \label{app:divergence-rho}
\end{equation}
The symbol $C$ denotes only this determinant constant, not $g_{xx}(0)$.

\subsection{Curvature and the gradient inequalities}

\begin{lemma}[Curvature and its full evolution]
\label{curvature}
Under Eq.~\eqref{app:constant-density},
\begin{align}
 R&=\frac{(\partial_x^2 g_{xx})}{g_{xx}^2}
       -\frac{2(\partial_x g_{xx})^2}{g_{xx}^3}
     =-\partial_x^2(g_{xx}^{-1}),\label{app:curvature}\\
 \partial_tR
 &=\frac{K}{4\gamma_g}\partial_x^2
                      \left(\frac{(\partial_x m)^2}{g_{xx}^2}\right),
 \label{app:curvature-flux}\\
 \partial_t(g_{xx}^2R)
 &=\frac{K}{4\gamma_g}\bigg[
      \partial_x^2[(\partial_x m)^2]\notag\\
 &\quad -4\frac{(\partial_x g_{xx})}{g_{xx}}\partial_x[(\partial_x m)^2]\notag\\
 &\quad +2(\partial_x m)^2\left(\frac{(\partial_x g_{xx})}{g_{xx}}\right)^2\bigg].
 \label{app:weighted-curvature}
\end{align}
\end{lemma}

\begin{proof}
For the diagonal metric, direct computation gives
\[
 R=\frac{g_{\theta\theta}(\partial_x g_{\theta\theta})(\partial_x g_{xx})
             -2g_{\theta\theta}g_{xx}(\partial_x^2 g_{\theta\theta})
             +g_{xx}(\partial_x g_{\theta\theta})^2}
            {2g_{\theta\theta}^2g_{xx}^2}.
\]
Substitute
\[
\begin{aligned}
 (\partial_x g_{\theta\theta})&=-\frac{C(\partial_x g_{xx})}{g_{xx}^2},\\
 (\partial_x^2 g_{\theta\theta})&=\frac{2C(\partial_x g_{xx})^2}{g_{xx}^3}
                         -\frac{C(\partial_x^2 g_{xx})}{g_{xx}^2}
\end{aligned}
\]
to obtain Eq.~\eqref{app:curvature}. Next,
\[
 \partial_t(g_{xx}^{-1})
 =-\frac{K}{4\gamma_g}\frac{(\partial_x m)^2}{g_{xx}^2},
\]
which proves Eq.~\eqref{app:curvature-flux}. Expanding its two spatial derivatives gives
\begin{equation}
\begin{aligned}
 \partial_tR
 &=\frac{K}{4\gamma_g}\bigg[
 \frac{\partial_x^2[(\partial_x m)^2]}{g_{xx}^2}
 -\frac{4(\partial_x g_{xx})\partial_x[(\partial_x m)^2]}{g_{xx}^3}\\
 &\quad -\frac{2(\partial_x m)^2(\partial_x^2 g_{xx})}{g_{xx}^3}
 +\frac{6(\partial_x m)^2(\partial_x g_{xx})^2}{g_{xx}^4}\bigg].
\end{aligned}
 \label{app:curvature-expanded}
\end{equation}
Multiply by $g_{xx}^2$, add $2g_{xx}(\partial_tg_{xx})R$, and use Eq.~ \eqref{app:curvature}. The terms containing $(\partial_x^2 g_{xx})$ cancel, leaving Eq.~\eqref{app:weighted-curvature}.
\end{proof}

\noindent At a spatial critical point of $(\partial_x m)^2$, the above is the exact identity
\begin{equation}
\begin{aligned}
 \partial_tR
 &=\frac{K}{4\gamma_g g_{xx}^2}\partial_x^2[(\partial_x m)^2]
   -\frac{K}{2\gamma_g}\frac{(\partial_x m)^2}{g_{xx}}R\\
 &\quad +\frac{K}{2\gamma_g}
                   \frac{(\partial_x m)^2(\partial_x g_{xx})^2}{g_{xx}^4}.
\end{aligned}
 \label{eq:Rmin}
\end{equation}
The last term is nonnegative and cannot be omitted in an upper bound. If a \emph{fixed} point $x_0$ remains a maximum of $(\partial_x m)^2$ and $(\partial_x g_{xx})(t,x_0)=0$ for all $t$, then
\begin{equation}
 R(t,x_0)\leq R(0,x_0)
                   \left(\frac{g_{xx}(0,x_0)}{g_{xx}(t,x_0)}\right)^2.
 \label{app:symmetric-curvature}
\end{equation}
Indeed Eq.~\eqref{app:weighted-curvature} has a nonpositive right-hand side at that fixed point. The exponent is $2$, and this is a one-sided estimate: it does not bound $|R|$ or imply $R\to0$.

\begin{lemma}[Coordinate gradient and its extrema]
\label{minimum}
The exact commuted equations are
\begin{align}
 \partial_t (\partial_x m)
 &=\frac{K}{\gamma_ m}\left[
      \frac{ (\partial_x^3 m)}{g_{xx}}
       -\frac{2(\partial_x g_{xx})}{g_{xx}^2} (\partial_x^2 m)-R (\partial_x m)\right],
 \label{app:commuted-rho}\\
 \partial_t(\partial_x m)^2
 &=\frac{K}{\gamma_ m g_{xx}}\partial_x^2[(\partial_x m)^2]
       -\frac{2K(\partial_x g_{xx})}{\gamma_ m g_{xx}^2}\partial_x[(\partial_x m)^2]
       \\
       &-\frac{2K}{\gamma_ m g_{xx}}(\partial_x^2 m)^2
       -\frac{2K}{\gamma_ m}R(\partial_x m)^2.
 \label{eq:curv}
\end{align}
At a nonzero interior maximum of $(\partial_x m)^2$,
\begin{equation}
 \partial_t(\partial_x m)^2\leq-\frac{2K}{\gamma_ m}R(\partial_x m)^2.
 \label{app:maximum-correct}
\end{equation}
At a strictly positive interior minimum the inequality is reversed.
\end{lemma}

\begin{proof}
Differentiate Eq.~\eqref{app:divergence-rho} in the fixed coordinate:
\[
\begin{aligned}
 \partial_t (\partial_x m)&=\frac{K}{\gamma_ m}
          \partial_x^2\left(\frac{\partial_x m}{g_{xx}}\right)
 \\
 &=\frac{K}{\gamma_ m}\bigg[
     \frac{ (\partial_x^3 m)}{g_{xx}}
     -\frac{2(\partial_x g_{xx}) (\partial_x^2 m)}{g_{xx}^2}\\
 &\quad +\bigg(\frac{2(\partial_x g_{xx})^2}{g_{xx}^3}
                -\frac{(\partial_x^2 g_{xx})}{g_{xx}^2}\bigg)(\partial_x m)\bigg].
\end{aligned}
\]
The last coefficient is $-R$. Multiplication by $2 (\partial_x m)$ and the identity $\partial_x^2[(\partial_x m)^2]=2(\partial_x^2 m)^2+2 (\partial_x m) (\partial_x^3 m)$ prove Eq.~ \eqref{eq:curv}. At a nonzero critical point, $ (\partial_x^2 m)=0$. The sign of $\partial_x^2[(\partial_x m)^2]$ then gives the asserted inequalities. A maximum only implies a nonpositive second derivative, not a strictly negative one.
\end{proof}

\noindent For a compact spatial domain with periodic or homogeneous Neumann conditions on $ m$, let $M(t)=\max_x( (\partial_x m)(t,x))^2$. The maximum is locally Lipschitz in time. At almost every time,
\begin{equation}
 \frac{dM}{dt}(t)\leq-\frac{2K}{\gamma_ m}M(t)
       \min_{\{x:( (\partial_x m)(t,x))^2=M(t)\}}R(t,x).
 \label{app:maximum-envelope}
\end{equation}
For $M(t)>0$ a Neumann boundary cannot be a maximizer, since $ (\partial_x m)=0$ there. To justify the envelope inequality, take a maximizer at time $t+h$ in the difference quotient, pass to an accumulation point as $h\downarrow0$, and apply Eq.~\eqref{app:maximum-correct}. This also gives the upper right Dini derivative version of Eq.~\eqref{app:maximum-envelope}. It supplies an upper bound for $M$, not the lower bound needed for pinching.

\noindent If a differentiable maximizer $x_m(t)$ is chosen and $R_m(t)=R(t,x_m(t))$, the chain rule gives
\begin{align}
 \frac{d}{dt}R_m(t)
   &=\partial_tR(t,x_m(t))\notag\\
   &\quad+\frac{dx_m}{dt}(t)(\partial_x R)(t,x_m(t)),\label{app:moving-R}\\
 \frac{d}{dt}g_{xx}(t,x_m(t))
   &=\frac{K}{4\gamma_g}M(t)
           +\frac{dx_m}{dt}(t)(\partial_x g_{xx})(t,x_m(t)).\label{app:moving-metric}
\end{align}
Only the derivative of $(\partial_x m)^2$ loses its transport term at a gradient maximum. Neither of these two equations is a closed ordinary differential inequality without additional hypotheses.

\subsection{Collapse to Pinching}
\label{sec:small-theorem}
We next derive criteria for transverse collapse.
\begin{proposition}
\label{app:exposure}
At each fixed spatial point, on every smooth existence interval,
\begin{align}
 g_{xx}(t,x)
 &=g_{xx}(0,x)\notag\\
 &\quad+\frac{K}{4\gamma_g}\int_0^t( (\partial_x m)(s,x))^2ds,
 \label{eq:final}\\
 l_{tr}(t,x)
 &=l_{tr}(0,x)\notag\\
 &\times\left(1+\frac{K}{4\gamma_g g_{xx}(0,x)}
                     \int_0^t( (\partial_x m)(s,x))^2ds\right)^{-1/2}
 \label{app:exact-circumference}
\end{align}
where $l_{tr}(t,x)=2\pi\sqrt{g_{\theta\theta}(t,x)}$ is the length of the transverse circle. For a global smooth solution,
\begin{equation}
 l_{tr}(t,x)\longrightarrow0
 \quad\Longleftrightarrow\quad
 \int_0^\infty( (\partial_x m)(s,x))^2ds=\infty.
 \label{app:collapse-iff}
\end{equation}
\end{proposition}

\begin{proof}
Integrate Eq.~\eqref{eq:gxx} at the fixed point and substitute into Eq.~\eqref{app:product}. Taking the square root gives Eq.~\eqref{app:exact-circumference}. Its denominator is monotone, which proves the equivalence. No evaluation along moving maxima is involved.
\end{proof}

\phantomsection
\noindent \emph{Collapse criterion at a general time.}
\label{app:general-time-collapse}
The radius identity, Eq.~\eqref{eq:exposure}, also gives a collapse criterion at a general terminal time. Suppose the solution is smooth for $0\leq t<T$, where $0<T\leq\infty$. The transverse circle at a fixed $x$ pinches, in the sense that $r(x,t)\to0$ as $t\to T^-$, if and only if

\begin{equation}
\int_0^T
(\partial_xm(x,s))^2\,ds=\infty.
\label{eq:collapse-criterion}
\end{equation}
The integral is understood as an improper limit as $t\to T^-$. This criterion concerns the evolving solution; it does not establish finite-time pinching or global existence from arbitrary initial data.

\noindent For $g_{xx}(0,x)=1$ and $g_{\theta\theta}(0,x)=r(0)^2$, the minimum radius is exactly
\begin{equation}
 r_{\min}(t)=r(0)
    \left(1+\frac K{4\gamma_g}
             \max_x\int_0^t( (\partial_x m)(s,x))^2ds\right)^{-1/2}.
 \label{app:min-radius}
\end{equation}
In general $\max_x\int_0^t(\partial_x m)^2ds$ cannot be replaced by $\int_0^t M(s)ds$; only the inequality in that direction is available. For noncompact $I$, replace the maximum by a supremum and the minimum by an infimum when they are not attained.

\begin{theorem}
\label{small}
Let a smooth global solution have a fixed point $x_0$ such that $ (\partial_x^2 m)(t,x_0)=0$ for all $t\geq0$. Suppose there is a nonnegative $\lambda\in L^1([0,\infty))$ for which
\begin{equation}
\begin{aligned}
 &R(t,x_0)( (\partial_x m)(t,x_0))^2\\
 &\quad-\frac{ (\partial_x m)(t,x_0) (\partial_x^3 m)(t,x_0)}{g_{xx}(t,x_0)}\\
 &\qquad\leq\lambda(t)( (\partial_x m)(t,x_0))^2.
\end{aligned}
 \label{app:full-loss-assumption}
\end{equation}
Set $B=\frac{2K}{\gamma_ m}\int_0^\infty\lambda(s)ds$. If $( (\partial_x m)(0,x_0))^2>0$, then
\begin{align}
 ( (\partial_x m)(t,x_0))^2
  &\geq ( (\partial_x m)(0,x_0))^2\notag\\
  &\quad \times\exp\!\left(-\frac{2K}{\gamma_ m}\int_0^t\lambda(s)ds\right),
 \label{app:gradient-lower}\\
 g_{xx}(t,x_0)
  &\geq g_{xx}(0,x_0)
          +\frac{K}{4\gamma_g}( (\partial_x m)(0,x_0))^2e^{-B}t.
 \label{app:growth-lower}
\end{align}
In particular, for any specified $T>0$, the explicit condition
\begin{equation}
 ( (\partial_x m)(0,x_0))^2
    \geq\frac{4\gamma_g g_{xx}(0,x_0)}{KT}e^B
 \label{app:corrected-threshold}
\end{equation}
implies $g_{xx}(t,x_0)\geq g_{xx}(0,x_0)(1+t/T)$. This conclusion holds for every $\delta>0$ under the stated diffusion-and-curvature hypothesis.
\end{theorem}

\begin{proof}
Evaluate Eq.~\eqref{app:commuted-rho} at the fixed point and multiply by $2 (\partial_x m)$:
\[
\begin{aligned}
 \partial_t(\partial_x m)^2(t,x_0)
 &=-\frac{2K}{\gamma_ m}\bigg[
       R(\partial_x m)^2\\
 &\qquad-\frac{ (\partial_x m) (\partial_x^3 m)}{g_{xx}}\bigg](t,x_0)\\
 &\geq-\frac{2K}{\gamma_ m}\lambda(t)( (\partial_x m)(t,x_0))^2.
\end{aligned}
\]
The integrating factor proves Eq.~\eqref{app:gradient-lower}, including positivity for all finite times. Integrate Eq.~\eqref{eq:gxx} and bound the exponential below by $e^{-B}$ to obtain Eq.~\eqref{app:growth-lower}. The last assertion is immediate.
\end{proof}

\noindent At a fixed nonzero gradient maximum, the full loss appearing here is
\begin{equation}
 R(t,x_0)-\frac{ (\partial_x^3 m)(t,x_0)}{g_{xx}(t,x_0) (\partial_x m)(t,x_0)}.
 \label{app:effective-loss}
\end{equation}
The second term is nonnegative at a maximum and records the loss due to diffusion. Negative curvature alone does not control it. For example, if the positive part of Eq.~\eqref{app:effective-loss} is bounded by $A(1+t/T)^{-p}$ with $A\geq0$ and $p>1$, then one may take $B=2KAT/(\gamma_ m(p-1))$. Such a bound must be proved for the full solution; it cannot be obtained by differentiating an assumed lower bound for $g_{xx}$. A growth estimate for $g_{xx}$ does not determine its derivative or $(\partial_x m)^2$.

\begin{corollary}
\label{transverse}
Under the assumptions of Theorem~\ref{small},
\begin{equation}
 l_{tr}(t,x_0)
 \leq l_{tr}(0,x_0)
 \left(1+\frac{K( (\partial_x m)(0,x_0))^2e^{-B}}
                       {4\gamma_g g_{xx}(0,x_0)}t\right)^{-1/2}.
 \label{app:transverse-rate}
\end{equation}
If Eq.~\eqref{app:corrected-threshold} holds, this is at most $l_{tr}(0,x_0)(1+t/T)^{-1/2}$.
\end{corollary}

\begin{proof}
Use Eqs.~\eqref{app:exact-circumference} and \eqref{app:gradient-lower}. The conclusion concerns the circle at the same fixed point $x_0$.
\end{proof}

\begin{proposition}
\label{app:bochner-comparison}
For periodic or homogeneous Neumann data, or bounded decaying data on $\mathbb R$, suppose $R(t,x)\geq-\kappa(t)$ on a smooth existence interval, where $\kappa\geq0$ is locally integrable. Put
\[
 M_0=\sup_x\frac{( (\partial_x m)(0,x))^2}{g_{xx}(0,x)},\qquad
 A(t)=\frac K{\gamma_ m}\int_0^t\kappa(s)ds,
\]
and define
\[
 D(t)=1+\frac K{4\gamma_g}M_0\int_0^t e^{A(s)}ds.
\]
Then
\begin{align}
 \sup_x\frac{( (\partial_x m)(t,x))^2}{g_{xx}(t,x)}
   &\leq\frac{M_0e^{A(t)}}{D(t)},\label{app:logistic}\\
 g_{xx}(t,x)&\leq g_{xx}(0,x)D(t),\label{app:metric-upper}\\
 l_{tr}(t,x)&\geq l_{tr}(0,x)D(t)^{-1/2}.
 \label{app:circumference-lower}
\end{align}
\end{proposition}

\begin{proof}
Set $M(t)=\sup_x|\nabla m(t,x)|_g^2$. The case $D_g=0$ of Eq.~\eqref{app:bochner} and the maximum principle imply
\[
 D^+M\leq\frac K{\gamma_ m}\kappa(t)M
                 -\frac K{4\gamma_g}M^2.
\]
Solving the comparison equation by substituting $M=e^{A(t)}/z$ yields Eq.~\eqref{app:logistic}; the case $M_0=0$ is identically zero. At a fixed $x$,
\[
\begin{aligned}
 \partial_t\log g_{xx}(t,x)
   &=\frac K{4\gamma_g}\frac{(\partial_x m)^2}{g_{xx}}\\
   &\leq\frac K{4\gamma_g}\frac{M_0e^{A(t)}}{D(t)}
   =\partial_t\log D(t).
\end{aligned}
\]
Integration and the product identity give the remaining bounds.
\end{proof}

\noindent For finite $t$ the denominator $D(t)$ is finite, so a locally integrable curvature lower bound prevents vanishing circumference on that smooth interval. This upper bound for metric growth supplies no lower growth estimate and hence no assertion of infinite-time pinching.

\subsection{Exact examples}
Maintaining different levels of morphogen activity at the tube ends prevents the axial gradient from disappearing everywhere and sustains transverse thinning, conditional on smooth existence. In the affine example below, the gradient is spatially uniform: the whole cylinder narrows and stretches, rather than developing a localized neck.
\begin{proposition}
\label{app:dirichlet-collapse}
On $[0,L]$, suppose $ m(t,0)=c_0$ and $ m(t,L)=c_L$ are fixed, with $c_L\ne c_0$, and the initial metric is $g_{xx}(0,x)=1$, $g_{\theta\theta}(0,x)=r(0)^2$. Every smooth solution satisfies
\begin{align}
 \frac1L\int_0^L g_{xx}(t,x)\,dx
 &\geq1+\frac K{4\gamma_g}
                      \left(\frac{c_L-c_0}{L}\right)^2t,
 \label{app:dirichlet-average}\\
 r_{\min}(t)
 &\leq r(0)\left[1+\frac K{4\gamma_g}
                      \left(\frac{c_L-c_0}{L}\right)^2t\right]^{-1/2}.
 \label{app:dirichlet-radius}
\end{align}
If the solution remains smooth for all time, its minimum radius tends to zero. This conclusion is valid for all relaxation-rate ratios. It concerns the minimum radius, without asserting that the shrinking point is fixed or coincides with a maximum of $(\partial_x m)^2$.
\end{proposition}

\begin{proof}
The boundary values imply $\int_0^L (\partial_x m)(t,x)\,dx=c_L-c_0$. Cauchy--Schwarz therefore gives
\[
 \int_0^L(\partial_x m)^2dx\geq\frac{(c_L-c_0)^2}{L}.
\]
Integrating Eq.~\eqref{eq:gxx} in space and then in time proves Eq.~\eqref{app:dirichlet-average}. Since the maximum is at least the average and $g_{\theta\theta}=r(0)^2/g_{xx}$,
\[
 r_{\min}(t)=\frac{r(0)}{\sqrt{\max_xg_{xx}(t,x)}},
\]
which proves Eq.~\eqref{app:dirichlet-radius}. The assertion at infinite time uses the stated global-existence hypothesis.
\end{proof}

\begin{proposition}[An exact solution]
\label{app:affine}
On $[0,L]$, prescribe the time-independent Dirichlet values $ m(t,0)=c_0$ and $ m(t,L)=c_0+qL$, with $q\ne0$. For constant positive initial metric coefficients and $ m(0,x)=c_0+qx$, the exact solution for every $\gamma_ m>0$ is
\begin{align}
  m(t,x)&=c_0+qx,\nonumber\\
 g_{xx}(t,x)&=g_{xx}(0)+\frac{Kq^2}{4\gamma_g}t,\nonumber\\
 g_{\theta\theta}(t,x)
 &=g_{\theta\theta}(0)
        \left(1+\frac{Kq^2}{4\gamma_g g_{xx}(0)}t\right)^{-1},
 \qquad R(t,x)=0.
 \label{app:affine-solution}
\end{align}
The transverse circumference decays exactly as $t^{-1/2}$.
\end{proposition}

\begin{proof}
The metric remains independent of $x$ and $ m$ is affine, so $\Delta m=0$. The first metric equation has constant right-hand side, and the second follows from conservation of their product. All equations and boundary values are satisfied; spatial constancy of the metric implies $R=0$.
\end{proof}

\noindent In the exactly frozen-morphogen limit $D_ m=0$, which is a different system from any finite $\gamma_ m$, one likewise obtains
\begin{align}
  m(t,x)&= m(0,x),\nonumber\\
 g_{xx}(t,x)&=g_{xx}(0,x)+\frac K{4\gamma_g}t( (\partial_x m)(0,x))^2,
 \label{app:frozen}\\
 l_{tr}(t,x)&=l_{tr}(0,x)
       \left(1+\frac{Kt( (\partial_x m)(0,x))^2}
                       {4\gamma_g g_{xx}(0,x)}\right)^{-1/2}.
 \label{app:frozen-circumference}
\end{align}
For constant $g_{xx}(0,x)=g_0>0$ and a fixed critical point $x_0$ of $ ( (\partial_x m)(0,x))^2$, the exact curvature is
\begin{equation}
 R(t,x_0)
 =\frac{\frac K{4\gamma_g}t\,
                    \partial_x^2(( (\partial_x m)(0,x))^2)|_{x=x_0}}
        {(g_0+\frac K{4\gamma_g}t( (\partial_x m)(0,x_0))^2)^2}.
 \label{app:frozen-curvature}
\end{equation}
At a nonzero gradient maximum this is nonpositive and tends to zero like $t^{-1}$ when the maximum is nondegenerate. If $H=( (\partial_x m)(0,x_0))^2>0$, its magnitude is bounded by $\bigl|\partial_x^2(( (\partial_x m)(0,\cdot))^2)(x_0)\bigr|/(4g_0H)$, since $\sup_{s\geq0}s/(g_0+sH)^2=1/(4g_0H)$.

\noindent This pointwise bound does not imply globally bounded curvature. At a simple zero of $ (\partial_x m)(0,x)$, the same formula gives
\begin{equation}
 R(t,x_0)=\frac{K}{2\gamma_g g_0^2}
                            t( (\partial_x^2 m)(0,x_0))^2,
 \label{app:frozen-curvature-growth}
\end{equation}
which grows linearly. For example $ m(0,x)=\sin x$ on a circle with $g_0=1$ gives $R(t,\pi/2)=Kt/(2\gamma_g)$, although the curvature at the shrinking neck $x=0$ tends to zero. The frozen formulas cannot be transferred to finite positive $D_ m$ by a continuity argument uniform in $t\in[0,\infty)$.

\phantomsection
\noindent \emph{Curvature at the Gaussian neck center.}
\label{app:gaussian-center-curvature}
In the main-text notation, the power-law example has a vanishing neck circumference and vanishing central curvature as $t\to\infty$. On the whole line, however, curvature grows without bound in the surrounding, outward-moving transition region, as shown below. Since $hr=r_0$, the appendix curvature formula Eq.~\eqref{eq:axisymmetric-curvature} becomes
\begin{equation}
R=-\frac{1}{r_0^2}\partial_x^2(r^2).
\label{eq:analytic-curvature}
\end{equation}
For the frozen Gaussian gradient profile, writing $b=\alpha t/(4a^2)$ gives
\begin{equation}
R(0,t)=-\frac{4b}{a^2(1+b)^2}\longrightarrow0
\qquad(t\to\infty).
\label{eq:neck-center-curvature}
\end{equation}

\noindent This central value does not give a global curvature bound.

\noindent For the Gaussian derivative in the main-text frozen example, $ (\partial_x m)(0,x)=a^{-1}e^{-x^2/a^2}$ on $\mathbb R$ with $g_0=1$, there is also a direct global obstruction. With $\alpha=K/\gamma_g$ and $x_t^2=\frac{a^2}{2}\log(\alpha t/(2a^2))$ for $t\geq2a^2/\alpha$, the exact metric has $\frac{\alpha t}{4a^2}e^{-2x_t^2/a^2}=\frac12$. Substitution into Eq.~\eqref{app:curvature} gives
\begin{equation}
 R(t,x_t)=\frac{16}{27a^2}\log\frac{\alpha t}{2a^2}
                       -\frac{8}{9a^2}.
 \label{app:gaussian-global-curvature}
\end{equation}
Although the central curvature tends to zero, the spatial curvature supremum on the whole line grows without bound as $t\to\infty$. The moving points $x_t$ eventually leave any fixed bounded interval.

\noindent Finally, $l_{tr}$ is the length of a transverse circle, which is a geodesic only when its transverse metric coefficient has zero spatial derivative. A vanishing circumference is an intrinsic geometric degeneration; it is not a topology change at a finite smooth time. For a compact surface with boundary the correct Gauss--Bonnet formula is
\begin{equation}
 \int_\Sigma R\,d\mu_g
       +2\int_{\partial\Sigma}k_g\,ds=4\pi\chi(\Sigma).
 \label{app:gauss-bonnet}
\end{equation}
Thus the zero Euler characteristic of a cylinder does not by itself make the total scalar curvature zero. On a periodic tube, which is a torus, the boundary term is absent, but the total curvature identity still does not determine curvature signs at gradient extrema. For realization as a surface of revolution in $\mathbb R^3$, the independent embedding condition is
\begin{equation}
 g_{xx}-\left(\partial_x\sqrt{g_{\theta\theta}}\right)^2\geq0
 \quad\Longleftrightarrow\quad
 g_{xx}g_{\theta\theta}\geq\frac14(\partial_x g_{\theta\theta})^2.
 \label{app:embedding-condition}
\end{equation}
The estimates in this appendix describe the intrinsic flow. They do not establish this extrinsic condition for arbitrary data, a uniform curvature bound for arbitrary collapsing solutions, or pinching for every nonzero initial gradient.

\section[Frozen fields and restricted geometric evolution]{Frozen fields and restricted geometric evolution}
\label{app:frozen-gradient-solution}
For a prescribed smooth field $m(x)$, the special case $f=0$ and $D_g=0$ yields, in an axisymmetric material frame,
\begin{align}
\partial_t(h^2)&=\frac{\Gamma_gK}{4}(m_x)^2,\\
\partial_t(r^2)&=-\frac{\Gamma_gK}{4}
\frac{r^2}{h^2}(m_x)^2.
\end{align}
For initially $h=1$ and $r=r_0$, one obtains
\begin{equation}
h^2=1+\frac{\Gamma_gK t}{4}(m_x)^2,
\qquad r=\frac{r_0}{h}.
\label{eq:frozen-exact}
\end{equation}
The radius is positive at every finite time and the area density $hr$ is fixed. This establishes a restricted exact solution, not stability or admissible Euclidean embeddability of arbitrary metric perturbations. In particular, prescribed activity cannot establish well-posedness of the genuinely coupled field--curvature system.

\section[Region size and scalar persistence]{Region size and scalar persistence}
\label{app:neck-width-diffusion}

\subsection{Gradient-driven contraction on the stretching metric}
\label{app:gradient-persistence}

We use only the gradient-driven full-metric example, with $f=D_g=0$, nonconserved Model A, no advection or sources, and initial $h=1$, $r=r_0$. Write $p_0(x)=\partial_xm(x,0)$ and assume that $p_0^2$ has a nonzero, nondegenerate local maximum at $x=0$. Then $P=p_0(0)\ne0$, $\partial_xp_0(0)=0$, and $P\partial_x^2p_0(0)<0$. The local expansion
\begin{equation}
p_0(x)^2=P^2\left[1-\frac{x^2}{\ell_0^2}+O(x^3)\right],
\qquad \ell_0^2=-\frac{P}{\partial_x^2p_0(0)},
\label{eq:neck-width}
\end{equation}
defines the initial-coordinate scale. It measures the variation of the driving profile in the initial coordinate, not the extent of a radius sublevel set. In particular, the region below the cutoff $\varepsilon r_0$ has zero width when an isolated minimum first reaches that cutoff.

\noindent The physical meridional length of a fixed band $|x|\leq\ell_0$ is
\begin{equation}
L_{\mathrm{neck}}(t)=\int_{-\ell_0}^{\ell_0}h(x,t)\,dx.
\label{eq:neck-physical-length}
\end{equation}

\noindent For $D_m=0$, define
\begin{equation}
\begin{gathered}
\tau_g=\frac{4}{\alpha P^2},\qquad u=\frac{t}{\tau_g},\\
H_0(x,t)=h_0(x,t)^2=1+\frac{\alpha t}{4}p_0(x)^2.
\end{gathered}
\end{equation}
Thus $H_0(0,t)=1+u$ and $T_\varepsilon=\tau_g(\varepsilon^{-2}-1)$. For the Gaussian $p_0=a^{-1}e^{-x^2/a^2}$, $\ell_0=a/\sqrt2$ and $\tau_g=4a^2/\alpha$. Its fixed-band physical length obeys
\begin{equation}
L_{\mathrm{neck}}(t)
=a\int_{-1/\sqrt2}^{1/\sqrt2}
\sqrt{1+u e^{-2y^2}}\,dy.
\end{equation}
Consequently, as $\varepsilon\to0$,
\begin{equation}
L_{\mathrm{neck}}(T_\varepsilon)
\sim\frac{a}{\varepsilon}
\int_{-1/\sqrt2}^{1/\sqrt2}e^{-y^2}\,dy.
\end{equation}
This is a proper meridional length, not an axial distance in an embedding space or a claim of continued embeddability.

\noindent For positive $D_m$, the exact coupled scalar equation is
\begin{equation}
\partial_tm=D_m\partial_x(H^{-1}\partial_xm),\qquad H=h^2.
\end{equation}
The local physical scale $h_0(0,t)\ell_0$ suggests the accumulated diffusion measure
\begin{equation}
\chi(t)=\frac{D_m}{\ell_0^2}\int_0^t\frac{ds}{H_0(0,s)}
=\eta\ln(1+u),\qquad
\eta=\frac{D_m\tau_g}{\ell_0^2}.
\label{eq:diffusion-clock-appendix}
\end{equation}
For short times this recovers $D_mt/\ell_0^2$. At the cutoff, $\chi(T_\varepsilon)=2\eta\ln(1/\varepsilon)$, whereas the initial-geometry estimate is $\eta(\varepsilon^{-2}-1)$.

\noindent The spatial dependence of $H$ must also be retained. Formally expanding about the frozen solution at fixed finite $u$, the first-order correction $\delta p$ to $p=\partial_xm$ satisfies
\begin{equation}
\partial_t\delta p
=D_m\partial_x^2(H_0^{-1}p_0),\qquad \delta p(x,0)=0.
\end{equation}
Changes of $H$ enter this scalar equation only at the next order, because its right-hand side already carries $D_m$. At $x=0$,
\begin{equation}
\frac{\partial_t\delta p(0,t)}{P}
=\frac{D_m}{\ell_0^2}\frac{u-1}{(1+u)^2}.
\end{equation}
The gradient initially weakens but its first-order rate changes sign at $u=1$: nonuniform diffusion on the stretching metric can subsequently steepen the coordinate gradient. Integrating gives
\begin{equation}
\frac{\delta p(0,t)}P
=\eta\left[\ln(1+u)-\frac{2u}{1+u}\right].
\label{eq:central-gradient-correction}
\end{equation}
Since $|u-1|/(1+u)^2\leq(1+u)^{-1}$, the magnitude of this first-order relative correction at every time up to $t$ is bounded by $\chi(t)$. Thus $\chi(T_\varepsilon)\ll1$ is a local perturbative consistency diagnostic that avoids cancellation in the signed correction. For the Gaussian, $\eta=8D_m/\alpha$, giving
\begin{equation}
\chi_\varepsilon
=\frac{16D_m}{\alpha}\ln\frac1\varepsilon.
\label{eq:neck-diffusion-clock}
\end{equation}

\noindent This calculation assumes a smooth perturbation over a fixed finite time interval. It supplies no bound on the higher-order remainder, the outer field, or the full spatial profile; boundary conditions and additional spatial scales can matter. It does not justify exchanging $D_m\to0$ with $\varepsilon\to0$, nor establish global existence or collapse for a freely diffusing morphogen. The frozen background and $T_\varepsilon$ are reference quantities, not the exact metric or cutoff time at positive $D_m$.

\subsection{Curvature-driven region and scalar relaxation}
\label{app:exponential-persistence}

We retain the full-metric curvature-coupling example with $f(m)=cm$, $c>0$, $K>0$, $D_g=0$, no advection or sources, and initial cylinder $h=1$, $r=r_0$. Let $\bar m(x)=m_0e^{-x^2/(2a^2)}$, with $m_0,a>0$, denote the reference frozen morphogen profile. On the initial cylinder, the curvature contribution to $-\partial_t\ln r$ is proportional to
\begin{equation}
q_e(x)=-\partial_x^2f(\bar m)
=\frac{cm_0}{a^2}\left(1-\frac{x^2}{a^2}\right)e^{-x^2/(2a^2)}.
\end{equation}
It is positive on $|x|<a$. Defining its local width by $q_e(x)=q_e(0)[1-x^2/\ell_e^2+O(x^4)]$ gives
\begin{equation}
\ell_e^2=-\frac{2\partial_x^2f(\bar m)(0)}
{\partial_x^4f(\bar m)(0)}=\frac{2a^2}{3}.
\label{eq:exponential-region-width}
\end{equation}
The derivative definition applies when $q_e$ has a nonzero, nondegenerate maximum at the morphogen peak. It characterizes the curvature contribution, not the total contraction rate when the gradient term is also present. In particular, it imposes no new restriction on $K$. A finite band's physical length is still $\int h\,dx$; $h(0,t)=1$ alone does not determine it or the later radius profile. These statements do not locate a global neck minimum or a radius-threshold boundary.

\noindent The frozen central law gives $T_\varepsilon=\tau_e\ln(1/\varepsilon)$ with $\tau_e=2a^2/(\Gamma_gcm_0)$ on smooth existence intervals. Restoring scalar mobility in the same Model A gives the full law
\begin{equation}
\partial_tm=D_m\Delta_gm-\Gamma_m cR,\qquad D_m=K\Gamma_m,
\label{eq:exponential-scalar-relaxation}
\end{equation}
where
\begin{equation}
\Delta_gm=\frac1{h^2}\left[
\partial_x^2m+
\left(\frac{\partial_xr}{r}-\frac{\partial_xh}{h}\right)
\partial_xm\right].
\end{equation}
The curvature term cannot be omitted while varying $\Gamma_m$ within this model. On the symmetric frozen background, denoted by $\bar g$, $\bar h(0,t)=1$ and $\Delta_{\bar g}\bar m(0)=-m_0/a^2$. Hence the formal first-order relative change of the peak value is
\begin{equation}
\frac{\delta m(0,t)}{m_0}
=-\frac{D_mt}{a^2}
-\frac{\Gamma_m c}{m_0}\int_0^t\bar R(0,s)\,ds.
\label{eq:exponential-peak-response}
\end{equation}
For $D_m>0$, the reference diffusion time across the morphogen profile's width $a$ is $\tau_{D,e}=a^2/D_m$. The diffusion contribution alone is therefore measured by
\begin{equation}
\chi_{e,D}=\frac{D_mT_\varepsilon}{a^2}
=\frac{2D_m}{\Gamma_g c m_0}\ln\frac1\varepsilon.
\label{eq:exponential-diffusion-comparison}
\end{equation}
The scale $a$ here is the width of the Gaussian morphogen profile; using $\ell_e$ instead would multiply this timescale ratio by $3/2$. A cancellation-free measure of the curvature contribution is
\begin{equation}
\chi_{e,R}(t)=\frac{\Gamma_m c}{m_0}
\int_0^t|\bar R(0,s)|\,ds.
\end{equation}
Small $\chi_{e,D}$ and $\chi_{e,R}(T_\varepsilon)$ bound the first-order relative change of the central morphogen activity. They do not control the higher-order remainder or the spatial derivatives that set the contraction rate. For a symmetric evolving morphogen profile, $h(0,t)=1$ and $\partial_t\ln r(0,t)=\Gamma_gc\partial_x^2m(0,t)/2$ continue to hold on smooth intervals; maintaining the exponential rate also requires persistence of $\partial_x^2m(0,t)$. Its first-order equation involves metric gradients and $\partial_x^2\bar R$, so a diffusion-time comparison alone is insufficient. No full-profile approximation, global existence, or freely relaxing collapse result follows from these local tests. Restoring scalar diffusion does not remove the short-wavelength instability of these $D_g=0$ curvature-coupled equations; see Appendix~\ref{app:coupled-stability}.

\section[Validity of the curvature-coupled model]{Validity of the curvature-coupled model}
\label{app:closure-validity}
The curvature-coupled model requires additional checks such as a well-posed time evolution before it can serve as a general evolution model.

\subsection{Frozen-field short-wavelength instability}
\label{app:frozen-stability}
For frozen $m$ and $D_g=0$, write $u=\log h$ and $v=\log r$. Linearizing the axisymmetric metric law about a smooth background gives the principal terms
\begin{equation}
\begin{aligned}
\partial_t\delta u&=a\,\partial_x\delta v+\text{l.o.t.},\\
\partial_t\delta v&=-a\,\partial_x\delta u+\text{l.o.t.},
\qquad a=\frac{\Gamma_g\partial_xf}{2h^2},
\end{aligned}
\label{eq:frozen-principal-validity}
\end{equation}
where l.o.t. denotes lower-order terms. Freezing the coefficients gives Fourier growth rates $\lambda_\pm=\pm ak$. Wherever $\partial_xf\ne0$, growth is unbounded as $|k|\to\infty$, obstructing standard Sobolev well-posedness. The exact exponential identity at a field peak, where $a=0$, therefore does not establish smooth evolution of the surrounding neck.

\subsection{Area balance}
\label{app:closed-area-balance}
Geometric homeostasis also requires an area balance. On a closed surface, Eq.~\eqref{eq:area-evolution} and Gauss--Bonnet give
\begin{equation}
\frac{d}{dt}\operatorname{Area}(\Sigma)
=-4\pi D_g\chi(\Sigma).
\label{eq:area-balance-validity}
\end{equation}
The unnormalized law therefore decreases spherical area at rate $8\pi D_g$ when $D_g>0$. Curvature smoothing alone does not maintain a stationary spherical area: area normalization, mechanical constraints, or growth sources would be needed. Convergence of normalized shape and preservation of physical area are distinct requirements.

\subsection{Dependence on the geometric response}
\label{app:response-dependence}
The contraction exponent likewise depends on the response law. For example, the radius-only law, Eq.~\eqref{eq:radius-smoothing-main}, with $h=1$, frozen $m$, and $f=D_g=0$, gives
\begin{equation}
r^2(x,t)=r_0^2-\Gamma_rK[\partial_xm(x)]^2t.
\label{eq:radius-law-validity}
\end{equation}
This differs from the full-metric power law. Both follow from specified mobilities and constraints, and remain subject to positivity and embedding. Neither specifies the physics of scission or the evolution after a singular neck forms.

Coupling Ricci flow to scalar or map evolution has an established precedent in harmonic--Ricci flow~\cite{Muller2012}. Its stress structure differs from the trace-free gradient contribution and the $Rf(m)$ coupling used here, so its existence results do not directly justify Eq.~\eqref{eq:metric-option-main}.

\section[Short-wavelength obstruction for the coupled Model A law]{Short-wavelength obstruction for the coupled Model A law}
\label{app:coupled-stability}
The nonnegative mobilities are not sufficient for well-posedness of the coupled equations when the curvature coupling is active. To see the obstruction, linearize around the flat cylindrical state $ds^2=dx^2+r_0^2d\theta^2$, $m=m_0$, where $R=0$, and set
\begin{equation}
m=m_0+\phi,\qquad r=r_0(1+\eta),\qquad c=f'(m_0).
\end{equation}
In the axisymmetric sector, to linear order,
\begin{equation}
\delta R=-2\partial_x^2\eta,\qquad
\delta(\Delta_gm)=\partial_x^2\phi,
\qquad
\delta\mathcal T_{\theta\theta}
=-2r_0^2c\,\partial_x^2\phi.
\end{equation}
The principal linearized subsystem of Model A coupled to Eq.~\eqref{eq:metric-option-main} is therefore
\begin{equation}
\partial_t\begin{pmatrix}\phi\\ \eta\end{pmatrix}
=\begin{pmatrix}
\Gamma_mK&2\Gamma_m c\\
\Gamma_gc/2&D_g
\end{pmatrix}
\partial_x^2\begin{pmatrix}\phi\\ \eta\end{pmatrix}.
\label{eq:linearized-system}
\end{equation}
Its diffusion matrix has determinant $\Gamma_m(KD_g-\Gamma_gc^2)$. For $\Gamma_m>0$, $\Gamma_g>0$, and $KD_g<\Gamma_gc^2$, the determinant is negative; hence one eigenvalue is negative. Fourier modes with wavenumber $k$ then exhibit growth $e^{\gamma k^2t}$, $\gamma>0$, which violates standard continuous dependence in Sobolev spaces. The strict inequality
\begin{equation}
KD_g>\Gamma_gc^2
\label{eq:necessary-principal-condition}
\end{equation}
is necessary for a strictly forward-parabolic principal matrix in this particular two-component axisymmetric sector. It is not sufficient to establish well-posedness of the full geometric system, which also requires control of gauge, other modes, possible boundaries, positivity of the metric, and embedding constraints. In particular, if $D_g=0$ and $\Gamma_gc\ne0$, the above coupled law is short-wavelength ill-posed near the flat state. In contrast, the exactly solvable example in Appendix~\ref{app:frozen-gradient-solution} has $f=0$ and a frozen field and does not exhibit this coupling instability.

\section[Embedding]{Embedding}
\label{app:embedding}

To realize an axisymmetric metric as a surface of revolution in $\mathbb R^3$, we use cylindrical coordinates ($r,\theta, z)$, where the radius of the tube $r(x,t) \equiv \sqrt{g_{\theta\theta}(x,t)}$ and $z$ is the coordinate along the axis of symmetry. By Pythagoras' theorem, $z$ is related to $x$ and $r$ as
\beq g_{xx}dx^2 = dz^2 + dr^2,\eeq
leading to
\begingroup
\beq
\begin{aligned}
z &= \int dx \sqrt{g_{xx} - (\partial_xr)^2}\\
  &= \int dx \frac{1}{\sqrt{g_{\theta\theta}}}\sqrt{\det g - \frac{1}{4}(\partial_xg_{\theta\theta})^2} .
\end{aligned}
\eeq
\endgroup
This surface-of-revolution realization requires a nonnegative argument under the square root, i.e., $\det g - \frac{1}{4}(\partial_xg_{\theta\theta})^2 \ge 0$. In the main-text notation, this is
\begin{equation}
h^2\geq(\partial_xr)^2.
\label{eq:embedding-condition}
\end{equation}

Physically, the available meridional distance must accommodate radial variation. The intrinsic evolution does not enforce this condition. The frozen Gaussian-gradient solution eventually violates this condition. Even when the condition holds, it establishes geometric realizability, not a three-dimensional mechanical evolution law.
\par

The scalar curvature $R$ is
\begingroup
\beq R = \frac{g_{\theta\theta}(\partial_xg_{\theta\theta})(\partial_xg_{xx})- 2 g_{\theta\theta} g_{xx}\partial_x^2g_{\theta\theta} + g_{xx}(\partial_xg_{\theta\theta})^2}{2 g_{\theta\theta}^2 g_{xx}^2} \eeq
\endgroup
which in terms of $g_{xx} = h(x,t)^2$ and $g_{\theta\theta} = r(x,t)^2$ becomes
\begingroup
\beq R = \frac{2[(\partial_xh)(\partial_xr) - h\partial_x^2r]}{h^3 r}.\eeq
\endgroup
We also assume azimuthal symmetry for $m$, i.e., $m(x,\theta,t) = m(x,t)$.

\section[Material transport and conservation]{Material transport and conservation}
\label{app:transport-alternatives}
Let $v^i$ denote the material velocity relative to the chosen coordinates. The Lie derivative of the metric satisfies $({\cal L}_v g)_{ij}=\nabla_iv_j+\nabla_jv_i$, and
\begin{equation}
\frac{1}{\sqrt g}\partial_t\sqrt g
=\frac12g^{ij}\partial_tg_{ij},\qquad
\nabla_iv^i=\frac{1}{\sqrt g}\partial_i(\sqrt g v^i).
\label{eq:jacobian-identities}
\end{equation}
It follows by direct expansion that
\begin{equation}
D_tm+m\Theta=\frac1{\sqrt g}\left[
\partial_t(\sqrt g\,m)+\partial_i(\sqrt g\,mv^i)
\right].\label{eq:transport-identity}
\end{equation}
Writing the diffusive flux as $J^i=-M\nabla^i\mu$, Eq.~\eqref{eq:density-option-main} becomes the exact local continuity equation
\begin{equation}
\partial_t(\sqrt g\,m)
+\partial_i[\sqrt g(mv^i+J^i)]
=\sqrt g S_B.
\end{equation}
For a fixed coordinate domain $\Omega$, the divergence theorem therefore gives
\begin{equation}
\frac{d}{dt}\int_\Omega mdA
=\int_\Omega S_BdA
-\int_{\partial\Omega}(mv^i+J^i)\nu_i\,ds_g.
\label{eq:mass-balance}
\end{equation}
For a material domain whose boundary moves with $v$ in these coordinates, the $mv$ contribution cancels against the moving boundary, leaving only the relative flux $J^i\nu_i$. Thus the conserved quantity is the total amount, not the pointwise concentration. The $m\Theta$ term is indispensable when the material area changes. For Model A, $D_tm=-\Gamma_m\mu+S_A$ is instead a kinetic assumption; there is no corresponding general conservation law.

\section[Transport and alternative dynamics]{Transport and alternative dynamics}
\label{app:general-transport-laws}

The examples in Section~\ref{sec:axisymmetric-dynamics} select particular responses in Eq.~\eqref{eq:general-dynamics}. Here we detail the alternative laws summarized in Fig.~\ref{fig:dynamics-map}, including the distinction between a nonconserved activity and an areal density, as well as geometric smoothing. We also discuss radius relaxation.

\emph{Transport and coordinate convention.} Flow terms describe transport and any specified coupling of geometry to motion. If $m$ is carried by $v^i$ relative to the chosen coordinates, $\mathcal A^{\mathrm{flow}}=-v^i\nabla_i m$; the geometric flow term requires its own kinematic convention. Flow can itself be generated by passive forces or active driving, so the flow label identifies its transport role, not its energetic origin. The velocity must be prescribed or determined by an additional mechanical law.
Let $v^i$ be the tangential material velocity relative to the surface coordinates. Scalar transport, metric deformation, and the fractional rate of area change are described by

\begin{equation}
\begin{aligned}
D_tm&=\partial_tm+v^i\nabla_i m,\\
\mathring g_{ij}
&=\partial_tg_{ij}
+\nabla_iv_j+\nabla_jv_i,\\
\Theta&=\frac12g^{ij}\mathring g_{ij}
=\partial_t\ln\sqrt g+\nabla_iv^i .
\end{aligned}
\label{eq:material-rates}
\end{equation}
The velocity terms account for motion relative to the coordinates; in $\mathring g_{ij}$ they form a Lie derivative. These material rates transform covariantly under time-dependent coordinate changes.

\emph{Scalar dynamics.---} The energetic derivative $\mu$ does not determine whether $m$ is an activity or a conserved areal concentration. Using the material-area rate $\Theta$ defined in Eq.~\eqref{eq:material-rates}, two choices combining passive relaxation, advection, and specified sources, with no additional active response and nonnegative mobilities $\Gamma_m$ and $M$, are
\begin{subequations}\label{eq:scalar-options-main}
\begin{align}
\partial_tm+v^i\nabla_i m&=-\Gamma_m\mu+S_A,
\label{eq:nonconserved-option-main}\\
\partial_tm+v^i\nabla_i m+m\Theta
&=\nabla_i(M\nabla^i\mu)+S_B.
\label{eq:density-option-main}
\end{align}
\end{subequations}
The left-hand sides are $D_tm$ and $D_tm+m\Theta$, respectively. The first law is nonconserved local relaxation (Model A); the second is a continuity law for an areal density (Model B), with flux $J^i=-M\nabla^i\mu$. Indeed, Eq.~\eqref{eq:density-option-main} is equivalent to
\begin{equation}
\partial_t(\sqrt g\,m)+
\partial_i\!\left[\sqrt g\,(mv^i+J^i)\right]
=\sqrt g\,S_B.
\label{eq:scalar-conservative-main}
\end{equation}
Thus $\int_\Sigma m\,dA_g$ is conserved on a closed surface when $S_B=0$; with a boundary, the flux relative to that boundary must vanish. The factor $m\Theta$ accounts for changes in material area. Model A requires no such dilution term, and passive relaxation with $S_A=0$ is not active driving. Appendix~\ref{app:transport-alternatives} gives the transport and boundary balances.

\emph{Geometric dynamics.---} A geometric response expressed in material rates can be written as

\begin{equation}
\mathring g_{ij}
=\mathcal R_{ij}+\mathcal A_{ij},
\label{eq:metric-dynamics}
\end{equation}
where $\mathcal R_{ij}$ is the mechanical response and $\mathcal A_{ij}$ describes imposed growth or active remodeling.

The metric response requires its own constitutive mobility and geometric admissibility conditions. For the nonconserved field, a possible intrinsic equation in the chosen coordinate frame is
\begin{equation}
\partial_t g_{ij}
=-\frac{\Gamma_g}{2}\mathcal T_{ij}-D_g Rg_{ij},
\qquad \Gamma_g,D_g\geq0,
\label{eq:metric-option-main}
\end{equation}
where $\mathcal T_{ij}=g_{ik}g_{j\ell}\mathcal T^{k\ell}$. Unlike Eq.~\eqref{eq:metric-dynamics}, this equation specifies $\partial_tg_{ij}$ rather than the material rate $\mathring g_{ij}$. The first term is gradient descent of $\mathcal F$ in the metric variables at fixed $m$; the second is an independently chosen two-dimensional Ricci-flow term. In material coordinates, pure Ricci flow satisfies $\partial_tR=D_g(\Delta_gR+R^2)$, so it does not imply monotone decay of curvature or of $\mathcal F$ (Appendix~\ref{app:ricci-flow}).

For Model B, the conjugate metric response under variations preserving the \emph{local} amount $\sqrt g\,m$ is instead
\begin{equation}
\mathcal T^{\mathrm B}_{ij}
=\mathcal T_{ij}-m\mu g_{ij}.
\label{eq:density-stress-main}
\end{equation}
The raised-index tensor is equivalently $\mathcal T_{\mathrm{dens}}^{ij}=\mathcal T^{ij}-m\mu g^{ij}$. Consequently, $-\Gamma_g\mathcal T^{\mathrm B}_{ij}/2$, not $-\Gamma_g\mathcal T_{ij}/2$, is the corresponding passive metric-gradient response for that choice of material variables. Using the fixed-$m$ tensor in a Model B evolution is a different phenomenological model and need not dissipate $\mathcal F$ (Appendix~\ref{app:metric-energy}). The activity examples retain metric variation at fixed $m$.

For Eq.~\eqref{eq:metric-option-main} with $V=Rf(m)$,
\begin{equation}
\Theta=\frac{\Gamma_g}{2}\Delta_gf(m)-D_gR + \nabla\cdot v.
\label{eq:area-evolution-main}
\end{equation}
On a closed surface, Gauss--Bonnet yields $d\operatorname{Area}_g/dt=-4\pi D_g\chi(\Sigma)$. Thus pure variational metric relaxation ($D_g=0$) preserves \emph{total} area in this model, but generally not local area; neither condition follows from conservation of scalar amount. Area constraints and imposed remodeling must be incorporated into the complete metric law (Appendix~\ref{app:area-evolution}). With advection and sources absent and $D_g=0$, the correctly paired Model A or Model B gradient flows dissipate $\mathcal F$. The Ricci term has no definite sign in the coupled energy balance, while nonnegative mobilities alone do not guarantee a well-posed coupled initial-value problem (Appendices~\ref{app:frozen-stability} and~\ref{app:coupled-stability}). The axisymmetric examples in Section~\ref{sec:axisymmetric-dynamics} employ explicitly stated constitutive and geometric assumptions.

On a fixed geometry, both passive scalar laws dissipate $\mathcal F$ when advection, sources, and boundary energy fluxes vanish. An evolving metric contributes mechanical work; active remodeling or sources can supply energy, so a stationary shape need not minimize $\mathcal F$. Appendix~\ref{app:metric-energy} gives the energy balances.

\emph{Radius relaxation.---} For the tube metric in Eq.~\eqref{eq:metric}, an alternative to full-metric evolution is to hold the meridional coefficient $h$ fixed and evolve only the transverse radius $r$. Variational radius relaxation can then be combined with a Ricci-inspired smoothing term, obtained by retaining the radius component of Ricci flow:
\begin{equation}
\begin{aligned}
\partial_tr
&=-\frac{\Gamma_r}{2\pi hr}
\left.\frac{\delta\mathcal F}{\delta r(x)}\right|_{h,m}
+\frac{D_g}{h}\partial_x\!\left(\frac{\partial_xr}{h}\right),\\
\partial_th&=0,
\end{aligned}
\label{eq:radius-smoothing-main}
\end{equation}
with $\Gamma_r,D_g\geq0$. This law changes the tube's radius while holding longitudinal distances fixed. When $h=1$, its smoothing term reduces to ordinary diffusion of the radius profile. It describes a different geometric response from the full-metric relaxation used in Section~\ref{sec:axisymmetric-dynamics}; Appendix~\ref{app:relaxation-details} gives the radius law and derives its force normalization and energy balance.

\section[Diffeomorphism identity and exact energy balances]{Diffeomorphism identity and exact energy balances}
\label{app:metric-energy}

For the general evolution in Eq.~\eqref{eq:general-dynamics}, let $\mathcal A=\mathcal A^{\mathrm{pass}}+\mathcal A^{\mathrm{act}}+\mathcal A^{\mathrm{flow}}$ and $\mathcal G_{ij}=\mathcal G_{ij}^{\mathrm{pass}}+\mathcal G_{ij}^{\mathrm{act}}+\mathcal G_{ij}^{\mathrm{flow}}$ denote the total constitutive responses in the chosen coordinate frame, excluding the separately specified sources. On a closed surface, differentiating $\mathcal F[m,g]$ with respect to time gives the exact energy balance

\beq
\frac{d\mathcal F}{dt}
=\int_\Sigma dA_g\left[
\mu(\mathcal A+S)
+\frac12\mathcal T^{ij}
(\mathcal G_{ij}+S_{ij})
\right].
\label{eq:configurational-power}
\eeq

A tangential infinitesimal diffeomorphism generated by $w^i$ produces $\delta m=w^i\nabla_i m$ and $\delta g_{ij}=\nabla_iw_j+\nabla_jw_i$. Applying the first-variation identity to a closed surface and integrating by parts, invariance of $\mathcal F$ gives
\begin{equation}
\nabla_i\mathcal T^{ij}=\mu\nabla^j m.
\label{eq:ward}
\end{equation}
Using this identity, Eq.~\eqref{eq:configurational-power} can also be written in terms of the material rates defined in Eq.~\eqref{eq:material-rates}: the added Lie-transport terms give a vanishing integrated diffeomorphism variation. Thus
\begin{equation}
\frac{d\mathcal F}{dt}
=\int_\Sigma\left(\mu D_tm
+\frac12\mathcal T^{ij}\mathring g_{ij}\right)dA.
\label{eq:material-power}
\end{equation}
For Model A coupled to $\mathring g_{ij}=-\Gamma_g\mathcal T_{ij}/2 -D_gRg_{ij}+S_{ij}$, and with scalar source $S_A$, this yields
\begin{eqnarray}
\frac{d\mathcal F}{dt}
={}&-\int_\Sigma\Gamma_m\mu^2dA
-\frac14\int_\Sigma\Gamma_g|\mathcal T|_g^2dA\nonumber\\
&+D_g\int_\Sigma R\Delta_gfdA
+\int_\Sigma\mu S_AdA\nonumber\\
&+\frac12\int_\Sigma\mathcal T^{ij}S_{ij}dA.
\label{eq:energy-A}
\end{eqnarray}
The first two terms are nonpositive; the Ricci term has no general sign.

For Model B, integration by parts gives
\begin{equation}
\int_\Sigma\mu\nabla_i(M\nabla^i\mu) dA
=-\int_\Sigma M|\nabla\mu|_g^2 dA,
\end{equation}
so that Eq.~\eqref{eq:material-power} becomes
\begin{eqnarray}
\frac{d\mathcal F}{dt}
={}&-\int_\Sigma M|\nabla\mu|_g^2 dA
+\int_\Sigma\mu S_B dA\nonumber\\
&+\frac12\int_\Sigma
\mathcal T_{\mathrm B}^{ij}\mathring g_{ij} dA.
\label{eq:power-B}
\end{eqnarray}
Using the Model B metric-gradient response $\mathring g_{ij}=-\Gamma_g\mathcal T^{\mathrm B}_{ij}/2 -D_gRg_{ij}+S_{ij}$ gives
\begin{eqnarray}
\frac{d\mathcal F}{dt}
={}&-\int_\Sigma M|\nabla\mu|_g^2dA
-\frac14\int_\Sigma
\Gamma_g|\mathcal T_{\mathrm B}|_g^2dA\nonumber\\
&+D_g\int_\Sigma R(\Delta_gf+m\mu)dA\nonumber\\
&+\int_\Sigma\mu S_BdA
+\frac12\int_\Sigma\mathcal T_{\mathrm B}^{ij}S_{ij}dA.
\label{eq:energy-B}
\end{eqnarray}
In particular, if $D_g=0$ and the sources vanish, both correctly paired systems are formally dissipative. If Model B is coupled instead to the original fixed-$m$ tensor $\mathcal T$, the geometric dissipation contribution is $-\frac14\int \Gamma_g\mathcal T_{\mathrm B}^{ij}\mathcal T_{ij}\,dA_g$, which does not have a definite sign. The use of $\mathcal T$ rather than $\mathcal T_{\mathrm B}$ then represents a separate constitutive choice, not passive gradient descent of the joint energy at fixed local amount. These balances require smooth fields and boundary conditions that eliminate surface-flux and variation terms.

\section[Ricci flow is an independent geometric mechanism]{Ricci flow is an independent geometric mechanism}
\label{app:ricci-flow}
In two dimensions $R_{ij}=\tfrac12Rg_{ij}$, so the unnormalized Ricci-flow term is $\partial_tg_{ij}=-D_gRg_{ij}$. With no other metric response, the curvature satisfies
\begin{equation}
\partial_tR=D_g(\Delta_gR+R^2),
\label{eq:curvature-flow}
\end{equation}
where the Laplacian has the convention $\Delta_g=\nabla^i\nabla_i$ and is nonpositive on Fourier modes. The leading part diffuses curvature, but the $R^2$ reaction shows that neither pointwise curvature magnitude nor the coupled energy $\mathcal F$ necessarily decreases. This flow is conformal, while the additional $\mathcal T_{ij}$ term generally has a nonzero traceless part.

\section[Ricci flow]{Ricci flow}
\label{app:ricci-background}

In two dimensions, Ricci flow preserves the conformal class and is a local differential law. It admits a separate variational formulation within that class~\cite{DubedatShen2022}, but need not decrease the coupled functional $\mathcal F$. Metric relaxation and Ricci flow can therefore coexist without their sum necessarily being a gradient descent of $\mathcal F$.

Ricci flow supplies a reference evolution for geometric smoothing. On closed surfaces, the normalized flow approaches constant curvature. The relevant convergence results are due to Hamilton and Chow~\cite{hamilton,chow}:

\begin{theorem}[\cite{hamilton}]
Let $(M,g)$ be a compact oriented Riemannian surface. (1) If $M$ is not diffeomorphic to the 2-sphere $\mathbb{S}^{2}$, then any metric g converges to a constant curvature metric under the normalized Ricci flow. (2) If $M$ is diffeomorphic to $\mathbb{S}^{2}$, then any metric $g$ with positive Gaussian curvature on $\mathbb{S}^{2}$ converges to a metric of constant curvature under the normalized Ricci flow.
\end{theorem}
 \begin{samepage}
Later Chow~\cite{chow} removed the assumption of initial positive Gauss curvature in $(2)$ and proved the following theorem:
\begin{theorem}[\cite{chow}]
If $g$ is any metric on $\mathbb{S}^{2}$, then under Hamilton's Ricci flow, the Gaussian curvature becomes positive in finite time, and therefore, by (2) in Hamilton's theorem, $g$ converges to a metric of constant curvature.
\end{theorem}
\end{samepage}
These results describe the approach to constant curvature under normalized Ricci flow. A simple local example explains why a thin cylindrical product metric supplies no intrinsic tendency to contract: each one-dimensional factor is flat, so the product has $R=0$. Its metric is therefore stationary under $\partial_tg_{ij}=-D_gRg_{ij}$, whatever the cylinder radius.

A coupled morphogen supplies an additional geometric response. In the model studied here, the morphogen gradients generate anisotropic deformation, while a spatially varying curvature coupling supplies a second route to constriction.

Figure~\ref{fig:pinching} illustrates four possible changes in connectivity. The local intrinsic calculation identifies a tendency toward constriction; its realization in any of these global scenarios requires matching to the surrounding surface and specifying the physics of scission or fusion.

\end{document}